\documentclass[10pt,journal,letterpaper,twocolumn,twoside]{IEEEtran} 

\usepackage{amsmath,amsfonts,bm,amssymb,psfrag,ifthen,color,subfigure}
\usepackage{mathrsfs}
\usepackage{stfloats}

\usepackage[justification=centering]{caption}

\usepackage[dvips]{epsfig}
\usepackage[dvips]{graphicx}
\usepackage{amsmath,amsfonts,bm,amssymb,psfrag,ifthen,color,subfigure}
\usepackage{algpseudocode}
\usepackage{algorithm}
\usepackage{algorithmicx}
\usepackage{setspace}
\usepackage{cite}
\usepackage{url}
\usepackage{longtable}
\usepackage{stfloats}  
\usepackage{graphics,booktabs,color,subfigure}
\usepackage{float}

\begin{document}%

\title{   Covert   Beamforming Design  for Intelligent Reflecting Surface  Assisted IoT  Networks}
  \author{Shuai Ma, Yunqi Zhang, Hang Li, Junchang Sun, Jia Shi,  Han Zhang, Chao Shen, and Shiyin Li.

\thanks{S. Ma, Y. Zhang, J. Sun,  and S. Li are with the School of Information and Control   Engineering, China
University of Mining and Technology, Xuzhou 221116,
China (e-mail: {mashuai001;ts19060151p31;sunjc;lishiyin}@cumt.edu.cn).}
\thanks{H. Li is with Data-driven Information System Laboratory, Shenzhen Research Institute of Big Data, Shenzhen 518172, Guangdong, China. (email: hangdavidli@163.com).}

\thanks{J. Shi is with the State Key Laboratory of ISN, School of Telecommunications Engineering, Xidian University, Xi'an 710071, China. (E-mail: jiashi@xidian.edu.cn).}
\thanks{H. Zhang is with Institute for Communication Systems, University of Surrey, Guildford GU2 7XH, U.K. (e-mail:han.zhang@surrey.ac.uk).}
\thanks{C. Shen is with the State Key Laboratory of Rail Traffic Control
and Safety, Beijing Jiaotong University, Beijing 100044, China (e-mail:
chaoshen@bjtu.edu.cn).}
}
\maketitle
\begin{abstract}
 In this paper, we consider
 covert   beamforming design  for
  intelligent reflecting surface (IRS)  assisted Internet of Things (IoT) networks, where  Alice  utilizes IRS to covertly transmit
  a message to Bob    without
being recognized by  Willie.
We investigate the joint beamformer design of Alice and IRS to maximize the covert rate of Bob when the knowledge about Willie's channel state information (WCSI) is perfect  and imperfect at Alice, respectively. For the former case, we develop a  covert beamformer  under the perfect covert constraint by applying semidefinite relaxation.
For  the later case,  the optimal   decision    threshold of  Willie is derived, and we analyze the false alarm and the missed detection probabilities.
Furthermore, we utilize the property of  Kullback-Leibler divergence to develop
the robust beamformer  based on a relaxation, S-Lemma and alternate iteration  approach.
Finally,   the numerical experiments evaluate the performance of the proposed covert beamformer design and robust beamformer design.

\end{abstract}

\IEEEpeerreviewmaketitle

\section{Introduction}


Internet of Things (IoT) has gradually been applied in various fields, e.g., industry, agriculture, and medicine.
 The number of smart communication devices increases tremendously, and data-hungry wireless applications also rapidly grow, resulting in the need of the high spectrum and energy efficiency \cite{Gong20Toward, Guo2020Energy}.
 Recently, intelligent reflecting surface (IRS) has  been considered as an effective solution which can improve the spectrum and energy-efficiency of the wireless networks by restructuring the wireless propagation environments.

IRS has drawn a wide attention for wireless communications applications.
Generally, the planar surface IRS consists of a large number of low-cost passive reflecting elements, each can reshape the phases, amplitudes, and reflecting angles of the incident signal independently \cite{QWu20Towards}, so that the propagation channel can be  intelligently adjusted to serve its own objectives.
Typically, by adaptively adjusting the phase shifts of the reflection elements, the signal reflected
by the IRS can be added constructively or destructively with the non-IRS-reflected signal to enhance the desired
signals or suppress the undesired signals \cite{Tan16Increasing}.
The advantages of the IRS aided IoT include low cost, low power consumption,  and simple construction. Moreover, the IRS  can improve  the received signal qualities by using its distinctive electromagnetic characteristics, such as negative refraction  \cite{GYu20Design}.

Owing to  the broadcast character of wireless communications, the IRS aided IoT is
    susceptible to
eavesdropping, especially  in
some public areas, e.g., airports, malls, and libraries.
Recently, many researches have investigated
optimization algorithms for improving the information security of  IRS aided IoT networks with respect to
  physical layer security \cite{Cui19Secure,Dong20Secure,XYu20Robust,Hong20Artificial,ZChu21Secrecy}.
   Physical layer security mainly focuses on preventing the transmitted wireless signal form from being decoded by the malicious users \cite{Barros11Physical}.
In  \cite{Cui19Secure}, the IRS was used to strengthen the desired signals and suppress the undesired signals for the secrecy rate maximization by  adaptively adjusting the phase shifts. In \cite{Dong20Secure}, the researchers studied an IRS assisted gaussian multiple-input multiple-output (MIMO) wiretap channel.
 In  \cite{XYu20Robust}, the authors investigated the joint design of the beamformers, artificial noise (AN) covariance matrix,  and the phase shifters at the IRSs, and the work considered the effect of the imperfect channel state information (CSI) of the eavesdropping channels.
  To explore the impact of the IRS on enhancing the security performance, the authors in \cite{Hong20Artificial}  proposed a block coordinate descent - Majorization Minimization (BCD-MM) algorithm for   AN-aided MIMO secure communication systems.
 In \cite{ZChu21Secrecy}, the closed-form expression of the secure precoder and the AN jamming precoder was derived by using the weighted
minimum mean square error (WMMSE) algorithm and Karush-Kuhn-Tucker (KKT) conditions, and the closed-form solution of the phase shift was obtained with the MM algorithm.

In fact, with growing security threats to the evolving wireless systems, though the transmitted
information is encrypted and the potential eavesdropping channel is physically restricted, the original data itself  might reveal confidential information. Covert communication aims to hide wireless signals from being discovered
by eavesdroppers. With the help of IRS,  \cite{Hossain20Intelligent} showed  that
  covert communication performance can be improved in the single-input-single-output (SISO) system.
   The work of \cite{Hossain20Intelligent} was then extended to a more general system setup
with both single antenna and multiple antennas at the legitimate transmitter in \cite{Si21arXiv}, and authors studied  the IRS-assisted covert communication  under the assumption of infinite number of channel uses. In \cite{Zhou20arXiv} authors considered the delay-constrained IRS assisted covert communication. In addition, the paper \cite{Wang21Intelligent} assumed that Bob generates jamming signals with a varying power to confuse Willie. More lately, the authors in \cite{Lv20arXiv} investigated covert communication in an IRS-assisted
non-orthogonal multiple access (NOMA) system.
In this work, we focus on a multiple-input-single-output (MISO) covert network with IRS.
Our key contributions are listed as follows:

\begin{itemize}

\item When
Willie's (eavesdropper)  channel state information (WCSI) is fully known at Alice, we consider maximizing the  covert rate of Bob (covert user)  under the quality of service (QoS) constraint of IRS, the covertness constraint, and the total power constraint.
To handle this non-convexity, the semidefinite relaxation (SDR) and the alternate iteration method are adopted.
We also evaluate the advantages of IRS by comparing with the system that is not aided by IRS.

\item For  the imperfect WCSI case,   the   optimal detection threshold   of Willie is  derived, and the corresponding detection error probability is obtained based on the robust beamformer vector. Such a result can be used  as a theoretical benchmark for evaluating the covert performance of the beamformers design.

\item Then, we further study   the joint  design of the robust beamformer and the IRS reflect beamformer with the objective of the achievable rate maximization,  subject to the perfect covert
transmission constraint, the total transmit power constraints of Alice and QoS of the IRS.
This non-convex problem is converted into a series of convex  subproblems with the methods of SDR,  S-lemma and the alternate iteration. Finally, the trade-off between the detection performance of Willie and the covert rate of Bob is illustrated in our simulation results.

 \end{itemize}

 The rest of this paper is given as follows. The
system model and the major assumptions are introduced in Section II.
A covert beamformer design with perfect WCSI is provided in Section III.
     The optimal   decision    threshold of  Willie  and the robust beamforming design with imperfect WCSI are presented in Section IV.
        In Section V, we evaluate the proposed beamformers based on
 the numerical results, and finally the paper
is concluded in Section   VI.

\emph{Notations}:  The vectors and matrices are represented by
boldfaced lowercase and uppercase letters, respectively.
The notations ${{\mathbb E}}\left\{  \cdot  \right\}$,
$\left\|  \cdot  \right\|$, ${\rm{Tr}}\left(  \cdot  \right)$, ${\mathop{\rm Re}\nolimits} \left(  \cdot  \right)$ and  ${\mathop{\rm Im}\nolimits} \left(  \cdot  \right)$ represent
the expectation,  Frobenius norm,  trace, the real part and imaginary part of its argument, respectively.
 The operator ${\bf{A}}\underline  \succ  {\bf{0}}$ means ${\bf{A}}$ is positive semidefinite.
The notation $\mathcal{CN}\left( {\mu ,{\sigma ^2}} \right)$ denotes a complex-valued circularly symmetric Gaussian distribution with   mean   $\mu$ and  variance   ${\sigma ^2}$.

\section{System Model}
\begin{figure}[h]
      \centering
	\includegraphics[width=8cm]{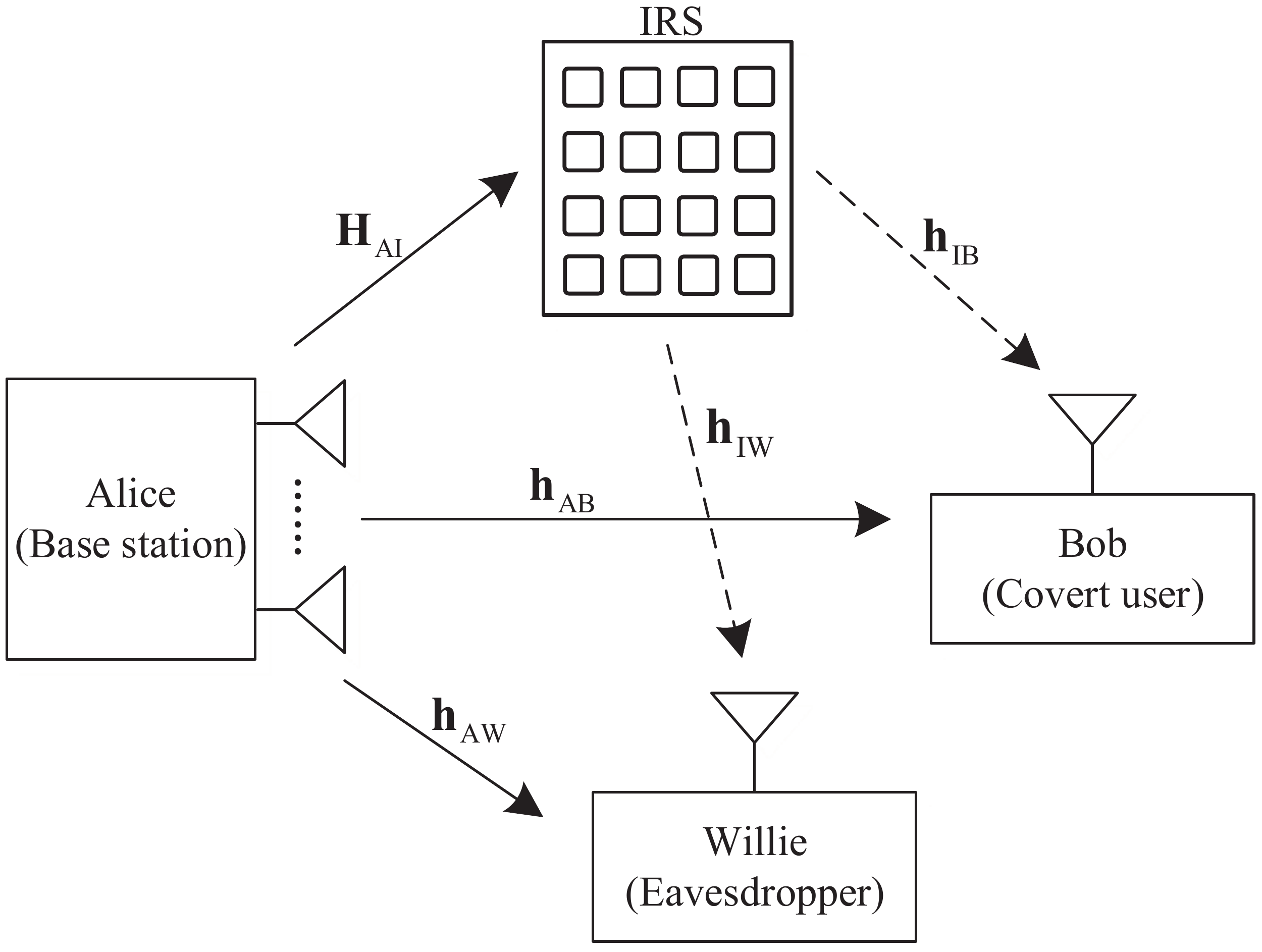}
 \caption{ Illustration of  the covert communication scenario.}
  \label{system} 
\end{figure}

A typical covert communication scenario is illustrated in Fig. 1, in which
Alice  transmits private data stream ${x_b}$  to Bob (covert user). For simplicity,  we assume  the power of the transmitted signal ${x_b}$ is unit, i.e., ${{\mathbb E}}\left\{ {{{\left| {{x_{\rm{b}}}} \right|}^2}} \right\} = 1$. At the same time, Willie (eavesdropper) silently (passively) supervise the communication environment and tries to identify whether Bob is receiving signals from Alice or not.
To safeguard  confidential signals
against eavesdropping, an IRS with a smart controller is adopted
to assist covert transmission, which is equipped with  $M$ reflecting units to coordinate the Alice and IRS for both channel acquisition and data transmission.
Specifically, the task for IRS is to adjust the phase shift coefficient of the reflection elements. Then, the signals from Alice is reflected passively to Bob from Alice-Bob link to constructively add with the non-reflected signal, and reflected passively to Willie from Alice-Willie link to destructively add with the non-reflected signal (without generating any extra noise).


In this model, we suppose that  Alice is equipped with $N$ antennas, while
    Bob and Willie each has a single antenna\footnote{
    Under this setup, Willie only needs to perform energy detection and does not have to know the beamforming vectors.}.
The channel coefficients from Alice to  Bob and Willie, from  the IRS to Bob and Willie are defined as ${{\bf{h}}_{\rm{AB}}} \in {\mathbb{C}^ {N \times 1}}$, ${{\bf{h}}_{\rm{AW}}} \in {\mathbb{C}^{N \times 1}}$,  ${{\bf{h}}_{\rm{IB}}} \in {\mathbb{C}^ {M \times 1}}$, and ${{\bf{h}}_{\rm{IW}}} \in {\mathbb{C}^{M \times 1}}$, respectively.
Besides, ${{\bf{H}}_{{\rm{AI}}}}\in {\mathbb{C}^ {M \times N}}$
  is the channel coefficients from Alice to the IRS. Here, ${\mathbb{C}^ {N \times M}}$
denotes a ${N \times M}$ complex-valued matrix.

In particular, the large-scale path loss is modeled as
${\rm{PL = }}\sqrt {{\zeta _0}{{\left( {{d_0}/d} \right)}^\alpha }} $ for all channels, where ${\zeta _0}$ is the path loss at the reference distance ${d_0} = 1{\rm{m}}$, $\alpha$ is the path loss exponent, and $d$ is the link distance.
For the small-scale fading, the channels from Alice  to  Willie and Bob, the is assumed to be Rayleigh fading, while the IRS-related
channels  is assumed to be Rician fading,
which are represented by
\begin{subequations}
\begin{align}
&{{\bf{H}}_{{\rm{AI}}}} = \left( {\sqrt {\frac{K_{{\rm{A}}}}{{1 + K}}} {\bf{H}}_{{\rm{A}}}^{{\rm{LOS}}} + \sqrt {\frac{1}{{1 + {K_{{\rm{A}}}}}}} {\bf{H}}_{{\rm{A}}}^{{\rm{NLOS}}}} \right),\\
&{{\bf{h}}_{{\rm{Ij}}}} = \left( {\sqrt {\frac{K_{{\rm{j}}}}{{1 + K}}} {\bf{h}}_{{\rm{j}}}^{{\rm{LOS}}} + \sqrt {\frac{1}{{1 + {K_{{\rm{j}}}}}}} {\bf{h}}_{{\rm{j}}}^{{\rm{NLOS}}}} \right),{\rm{j}} \in \left\{ {{\rm{B}},{\rm{W}}} \right\},
\end{align}
\end{subequations}
where ${K_{{\rm{j}}}}$ is the Rician factor, and ${\rm{j}}$ denotes ${\rm{B}}$, ${\rm{W}}$ and ${\rm{A}}$. ${\bf{H}}_{{\rm{A}}}^{{\rm{LOS}}}$ and ${\bf{H}}_{{\rm{A}}}^{{\rm{NLOS}}}$ are the deterministic  line-of-sight (LoS) and  the non-LoS (NLoS) of the Alice-IRS channel,  respectively. And the deterministic  LoS and  the NLoS of the IRS-related channel are represented by ${\bf{h}}_{{\rm{j}}}^{{\rm{LOS}}}$ and ${\bf{h}}_{{\rm{j}}}^{{\rm{NLOS}}}$,  respectively. Here, ${\bf{h}}_{{\rm{j}}}^{{\rm{LOS}}}$ is modeled as the product of steering vectors of the transmit and receive arrays, i.e., ${\bf{h}}_{{\rm{j}}}^{{\rm{LOS}}} = {{\bf{a}}_r}{\bf{a}}_t^H$, and  ${\bf{h}}_{{\rm{j}}}^{{\rm{NLOS}}}$  is modeled as Rayleigh fading. ${{\bf{a}}_r}$ and ${{\bf{a}}_t}$ are defined as,
\begin{small}
\begin{subequations}
\begin{align}
&{{\bf{a}}_r} = {\left[ {1,\exp \left( { \frac{{j2\pi{d_t}}}{\lambda }\sin {\varphi _t}} \right), \cdots ,\exp \left( { \frac{{j2\pi{d_t}}}{\lambda }\left( {{N_t} - 1} \right)\sin {\varphi _t}} \right)} \right]^H},\\
&{{\bf{a}}_t} = {\left[ {1,\exp \left( { \frac{{j2\pi{d_r}}}{\lambda }\sin {\varphi _r}} \right), \cdots ,\exp \left( { \frac{{j2\pi{d_r}}}{\lambda }\left( {{N_r} - 1} \right)\sin {\varphi _r}} \right)} \right]^H}.
\end{align}
\end{subequations}
\end{small}
Here, $\lambda$ denotes  the frequency of the carrier wave. ${d_t}$ and ${d_r}$ are the antenna spacing of the transmit and receive array, respectively. We assume that $\lambda  = \frac{{{d_r}}}{2} = \frac{{{d_t}}}{2}$. $\varphi _t$ and $\varphi _r$ are the angle of departure and the angle of arrival, where  ${\varphi _t} = {\tan ^{ - 1}}\left( {\frac{{{y_r} - {y_t}}}{{{x_r} - {x_t}}}} \right)$ and ${\varphi _r} = \pi  - {\varphi _t}$, with $\left( {{x_t},{y_t}} \right)$  and $\left( {{x_r},{y_r}} \right)$  being the location of the transmitter and the receiver, respectively.
The number of antennas at the transmitter and receiver are denoted as $N_t$ and $N_r$, respectively.

\subsection{Signal Model}
Let ${{\cal H} _0}$ denote the null hypothesis  that
 Alice does not transmit private  data stream to Bob, while 	${{\cal H} _1}$ denotes the alternate hypothesis that
Alice  transmits private data stream to  Bob\cite{Lehmann_2005_Testing}.
From Willie's perspective, Alice's transmitted signal   is given by
\begin{align}{\bf{x}} = \left\{ \begin{array}{l}
 0,~~~~~~~~{{\cal H} _0} \\
 {{\bf{w}}_{\rm{b}}}{x_{\rm{b}}},~~~{{\cal H} _1} \\
 \end{array} \right.
 \end{align}
where  ${{\bf{w}}_{\rm{b}}}$ denotes the  transmit beamformer vector for  ${{x_{{\rm{b}}}}}$. For simplicity, we also assume that Alice does not transmit any signal under ${{\cal H} _0}$.
The  beamformer  ${{\bf{w}}_{\rm{b}}}$ satisfies
\begin{align}
{\left\| {{{\bf{w}}_{\rm{b}}}} \right\|^2} \le {P_{{\rm{total}}}},
 \end{align}
where ${P_{{\rm{total}}}}$ denotes the    maximum transmit power   of Alice.

The diagonal matrix ${\bf{Q}} \buildrel \Delta \over = {\rm{diag}}\left( {\bf{q}} \right)$ is denoted as the phase shift matrix  at IRS, which diagonal elements are the corresponding elements of the vector ${\bf{q}}$.
Here,  ${\bf{q}} \buildrel \Delta \over = {\left[ {{q_1}, \cdots, {q_m}, \cdots, {q_M}} \right]^T}$ with ${q_m} = {\beta _m}{e^{j{\theta _m}}}$, where ${\theta _m} \in \left[ {0,2\pi } \right)$ and ${\beta _m} \in \left[ {0,1} \right]$ respectively denote the  controllable phase shift and amplitude
reflection coefficient,  introduced by the $m$th unit for $m = 1, \cdots M$.  For simplicity, we set ${\beta _m} = 1,\forall m$, to obtain the maximum gain of the reflecting power. Thus, we have
\begin{align}\label{qn}
\left| {{q_m}} \right| = 1,m = 1, \cdots M.
\end{align}

Thus, the received signal at Bob is expressed as
\begin{align}\label{yb}
{y_{{\rm{b}}}}=\left\{ {\begin{array}{*{20}{c}}
{z_{\rm{b}}},~~~~~~~~~~~~~~~~~~~~~~~~~~~~~~~~{{\cal H} _0}\\
\left( {{\bf{h}}_{{\rm{IB}}}^H{\bf{Q}}{{\bf{H}}_{{\rm{AI}}}} + {\bf{h}}_{{\rm{AB}}}^H} \right){{\bf{w}}_{\rm{b}}}{x_{\rm{b}}} + {z_{\rm{b}}},~{{\cal H} _1}
\end{array}} \right.
\end{align}
where ${z_{\rm{b}}}\sim {\cal CN}\left( {{\rm{0}},\sigma _{\rm{b}}^2} \right)$  is the received noise   at Bob.

For Willie, the received signal is given by
\begin{align}\label{yw}
{y_{{\rm{w}}}}=\left\{ {\begin{array}{*{20}{c}}
 {z_{\rm{w}}},~~~~~~~~~~~~~~~~~~~~~~~~~~~~~~~{{\cal H} _0}\\
 \left( {{\bf{h}}_{{\rm{IW}}}^H{\bf{Q}}{{\bf{H}}_{{\rm{AI}}}} + {\bf{h}}_{{\rm{AW}}}^H} \right){{\bf{w}}_{\rm{b}}}{x_{\rm{b}}} + {z_{\rm{w}}},~{{\cal H} _1}
\end{array}} \right.
\end{align}
where
${z_{\rm{w}}}\sim {\cal CN}\left( {{\rm{0}},\sigma _{\rm{w}}^2} \right)$    is the received noise   at Willie.

According to \eqref{yb}, the instantaneous rate at  Bob  under hypothesis ${{\cal H} _1}$ is  given by
\begin{align}\label{rateb}
{R_{\rm{b}}} = {\log _2}\left( {1 + \frac{{{{\left| {\left( {{\bf{h}}_{{\rm{IB}}}^H{\bf{Q}}{{\bf{H}}_{{\rm{AI}}}} + {\bf{h}}_{{\rm{AB}}}^H} \right){{\bf{w}}_{\rm{b}}}} \right|}^2}}}{{\sigma _{\rm{b}}^2}}} \right).
\end{align}

\subsection{Covert Constraints}

Since Willie needs to distinguish between  the two hypotheses ${{\cal H} _0}$ and ${{\cal H} _1}$ according to its received signal ${y_{{\rm{w}}}}$, it is necessary to quantify the probability of ${y_{{\rm{w}}}}$.
We assume that the likelihood functions of the received signals of Willie under ${{{\cal H}_{\rm{0}}}}$ and ${{{\cal H}_{\rm{1}}}}$ are expressed as ${p_0}\left( {{y_{\rm{w}}}} \right)$
 and ${p_1}\left( {{y_{\rm{w}}}} \right)$, respectively.
 Specifically, as per \eqref{yw}, ${p_0}\left( {{y_{\rm{w}}}} \right)$ and ${p_1}\left( {{y_{\rm{w}}}} \right)$ are given  by
\begin{subequations}\label{C_8}
\begin{align}
& {p_0}\left( {{y_{\rm{w}}}} \right) = \frac{1}{{\pi {\lambda _0}}}\exp \left( { - \frac{{{{\left| {{y_{\rm{w}}}} \right|}^2}}}{{{\lambda _0}}}} \right), \\
 &{p_1}\left( {{y_{\rm{w}}}} \right) = \frac{1}{{\pi {\lambda _1}}}\exp \left( { - \frac{{{{\left| {{y_{\rm{w}}}} \right|}^2}}}{{{\lambda _1}}}} \right),
\end{align}
  \end{subequations}
where ${\lambda _0} \buildrel \Delta \over = \sigma _{\rm{w}}^2$ and ${\lambda_1} \buildrel \Delta \over =
{\left| {\left( {{\bf{h}}_{{\rm{IW}}}^H{\bf{Q}}{{\bf{H}}_{{\rm{AI}}}} + {\bf{h}}_{{\rm{AW}}}^H} \right){{\bf{w}}_{\rm{b}}}} \right|^{\rm{2}}}+ \sigma _{\rm{w}}^2$.

In general,    the prior
probabilities of   hypotheses ${{{\cal H}_0}}$ and ${{{\cal H}_1}}$  are assumed to be equal, i.e., each equals to  $1/2$.
As such, the detection error probability is utilized to evaluate the detection performance, which can be given by \cite{Yan2019Gaussian,Cover_2003_Elements,Lehmann_2005_Testing}
\begin{align}\label{xi}
\xi  = \Pr \left( {{{\cal D}_1}\left| {{{\cal H}_0}} \right.} \right) + \Pr \left( {{{\cal D}_0}\left| {{{\cal H}_1}} \right.} \right) ,
\end{align}
where   ${{\cal D}_1}$ indicates  that the transmission from Alice to Bob is present, and  ${{\cal D}_0}$ indicates the other case.
In covert communications, Willie aims to minimize the probability of detection error $ {\xi } $ based on an optimal detector.
To be specific, the covert communication constraint
 can be written as  $ \xi  \ge 1 - \varepsilon $, where
$0 \le \varepsilon  \le 1$
is a priori value to specify the covert
communication constraint.

In order to incorporate $\xi$ into our problem formulation,  we next specify the conditions of the likelihood function so that with the given $\varepsilon$ the covert communication can be achieved (i.e., the constraint can be satisfied).
First, let
\begin{align}\label{C_7}
\xi
 = 1 - { V_T}\left( {{p_0},{p_1}} \right),
\end{align}
where ${V_T}\left( {{p_0},{p_1}} \right)$ is the total variation between ${p_0}\left( {{y_{\rm{w}}}} \right)$ and
${p_1}\left( {{y_{\rm{w}}}} \right)$.
Usually, computing ${V_T}\left( {{p_0},{p_1}} \right)$
analytically is intractable. Thus, we adopt  Pinsker's inequality \cite{Cover_2003_Elements}, and obtain
\begin{subequations}
\begin{align}
& { V_T}\left( {{p_0},{p_1}} \right) \le \sqrt {\frac{1}{2}D\left( {{p_0}\left\| {{p_1}} \right.} \right)},  \label{C_3}\\
 &{ V_T}\left( {{p_0},{p_1}} \right) \le \sqrt {\frac{1}{2}D\left( {{p_1}\left\| {{p_0}} \right.} \right)},\label{C_4}
\end{align}
  \end{subequations}
where $D\left( {{p_0}\left\| {{p_1}} \right.} \right)$ denotes the Kullback-Leibler (KL) divergence from $p_0(y_{\rm{w}})$ to $p_1(y_{\rm{w}})$, and
$D\left( {{p_1}\left\| {{p_0}} \right.} \right)$ is the KL divergence from $p_1(y_{\rm{w}})$ to $p_0(y_{\rm{w}})$.
$D\left( {{p_0}\left\| {{p_1}} \right.} \right)$   and
$D\left( {{p_1}\left\| {{p_0}} \right.} \right)$ are respectively given as

  \begin{small}
 \begin{subequations}
\begin{align}
  D\left( {{p_0}\left\| {{p_1}} \right.} \right) &= \int_{ - \infty }^{ + \infty } {{p_0}\left( {{y_{\text{w}}}} \right)\ln \frac{{{p_0}\left( {{y_{\text{w}}}} \right)}}{{{p_1}\left( {{y_{\text{w}}}} \right)}}} dy  = \ln \frac{{{\lambda _1}}}{{{\lambda _0}}} + \frac{{{\lambda _0}}}{{{\lambda _1}}} - 1, \label{C_5}\\
  D\left( {{p_1}\left\| {{p_0}} \right.} \right) &= \int_{ - \infty }^{ + \infty } {{p_1}\left( {{y_{\text{w}}}} \right)\ln \frac{{{p_1}\left( {{y_{\text{w}}}} \right)}}{{{p_0}\left( {{y_{\text{w}}}} \right)}}} dy = \ln \frac{{{\lambda _0}}}{{{\lambda _1}}} + \frac{{{\lambda _1}}}{{{\lambda _0}}} - 1. \label{C_6}
  \end{align}
\end{subequations}
\end{small}

Therefore,   in order to  achieve  covert communication with the given $\varepsilon$, i.e., $ \xi    \ge 1 - \varepsilon $, the KL divergences  of the likelihood functions should satisfy  one of the following constraints:
\begin{subequations}\label{Dp0p1}
\begin{align}
&D\left( {{p_0}\left\| {{p_1}} \right.} \right) \le 2{\varepsilon ^2},\\
&D\left( {{p_1}\left\| {{p_0}} \right.} \right) \le 2{\varepsilon ^2}.
\end{align}
\end{subequations}

\section{Proposed  Covert Transmission for Perfect WCSI}
Practically,
 we consider a condition that often appears. In this case,   Willie is a legitimate user.
 Alice knows the complete CSI of the channel ${{\bf{h}}_{\rm{IW}}}$ and ${{\bf{h}}_{\rm{AW}}}$,
  and then uses it to help Bob avoid Willie's monitoring \cite{Bash13,Forouzesh20Communication}.

Generally, in order to maximize the covert  rate to Bob  with given  vector
${{\bf{w}}_{\rm{b}}}$, the design of the IRS reflecting beamforming vector  ${\bf{q}}$
satisfies the following goal. That is, the phase of the reflected channel ${{\bf{h}}_{{\rm{IB}}}^H{\bf{Q}}{{\bf{H}}_{{\rm{AI}}}}}$ is aligned with that of the direct channel ${{{\bf{h}}_{{\rm{AB}}}}}$. In this way, we can maximize  the received signal power at
the user, which is equivalent to maximize ${R_{\rm{b}}}$.
Therefore,  we  maximize   the covert  rate to Bob by optimizing beamformers ${{\bf{w}}_{\rm{b}}}$ at Alice and the reflecting beamforming vector ${\bf{q}}$ at IRS.

 \subsection{Covert Beamformers  Design for Continuous Phase Shifts}
Specifically, we study a joint  beamforming  design problem with the objective of maximizing
  the achievable covert  rate of Bob  ${R_{\rm{b}}}$, subject to the   perfect covert transmission constraint, the total transmit power constraints of Alice and the IRS-related QoS. This is mathematically expressed as
\begin{subequations}\label{C_13}
\begin{align}
\mathop {\max }\limits_{{{\bf{w}}_{\rm{b}}},{\bf{q}}} {\rm{ }}&~{R_{\rm{b}}} \hfill \\
  {\text{s}}{\text{.t}}{\text{.}}~&D\left( {{p_0}\left\| {{p_1}} \right.} \right)= 0, \hfill \label{C_10}\\
 & {\left\| {{{\mathbf{w}}_{\rm{b}}}} \right\|^2} \le  {P_{{\rm{total}}}}, \hfill \\
 &\left| {{q_m}} \right| = 1,\forall m.
  \end{align}
\end{subequations}
  Note that problem \eqref{C_13} is   non-convex and  difficult to be optimally solved.
Moreover,      $D\left( {{p_0}\left\| {{p_1}} \right.} \right)=0$ or $D\left( {{p_1}\left\| {{p_0}} \right.} \right)=0$ means  that the  perfect covert transmission.

 Equivalently,
   problem \eqref{C_13} can be reformulated as
   \begin{subequations}\label{C_14}
\begin{align}
 \mathop {\max }\limits_{{{\mathbf{w}}_{\rm{b}}},{\bf{q}}} ~&{\left| {\left( {{\bf{h}}_{{\rm{IB}}}^H{\bf{Q}}{{\bf{H}}_{{\rm{AI}}}} + {\bf{h}}_{{\rm{AB}}}^H} \right){{\bf{w}}_{\rm{b}}}} \right|^2}\label{C_14a} \\
 {\rm{s.t.}}~&{\left| {\left( {{\bf{h}}_{{\rm{IW}}}^H{\bf{Q}}{{\bf{H}}_{{\rm{AI}}}} + {\bf{h}}_{{\rm{AW}}}^H} \right){{\bf{w}}_{\rm{b}}}} \right|^{\rm{2}}} = 0, \label{C_14b}\\
 & {\left\| {{{\mathbf{w}}_{\rm{b}}}} \right\|^2} \le  {P_{{\rm{total}}}}, \hfill \label{C_14c}\\
 &\left| {{q_m}} \right| = 1,\forall m.\label{C_14d}
 \end{align}
\end{subequations}

Next,  we need to solve the following two sub-problems iteratively: fix ${{\mathbf{w}}_{\rm{b}}}$ to optimize ${\bf{q}}$, and then fix ${\bf{q}}$ to optimize ${{\mathbf{w}}_{\rm{b}}}$,  which are given in the following two subsections in detail, respectively. Then, the entire algorithm is presented.

  \subsubsection{{{Sub-Problem 1. Optimizing ${{\mathbf{w}}_{\rm{b}}}$ with Given ${\bf{q}}$}}}

When we fix ${\bf{q}}$, problem \eqref{C_14} can be converted to the following problem
\begin{align}\label{C_20}
 \mathop {\max }\limits_{{{\mathbf{w}}_{\rm{b}}}} ~&{\left| {\left( {{\bf{h}}_{{\rm{IB}}}^H{\bf{Q}}{{\bf{H}}_{{\rm{AI}}}} + {\bf{h}}_{{\rm{AB}}}^H} \right){{\bf{w}}_{\rm{b}}}} \right|^2}\\
 {\rm{s.t.}}~&\eqref{C_14b},\eqref{C_14c}\notag.
 \end{align}

Let ${{\bf{t}}_{\rm{B}}} \buildrel \Delta \over = {\left( {{\bf{h}}_{{\rm{IB}}}^H{\bf{Q}}{{\bf{H}}_{{\rm{AI}}}} + {\bf{h}}_{{\rm{AB}}}^H} \right)}$ and ${{\bf{t}}_{\rm{W}}} \buildrel \Delta \over = \left( {{\bf{h}}_{{\rm{IW}}}^H{\bf{Q}}{{\bf{H}}_{{\rm{AI}}}} + {\bf{h}}_{{\rm{AW}}}^H} \right)$.
We  use the SDR technique, i.e., ${{\bf{W}}_{\rm{b}}}{\rm{ = }}{{\bf{w}}_{\rm{b}}}{\bf{w}}_{\rm{b}}^H$. By ignoring the rank-one constraints,  we obtain a relaxed version of problem \eqref{C_20}   as
    \begin{subequations}\label{problem1B}
\begin{align}
 \mathop {\max }\limits_{{{\bf{W}}_{\rm{b}}}} ~&{\rm{Tr}}\left( {{\bf{t}}_{\rm{B}}{{\bf{W}}_{\rm{b}}}{{\bf{t}}_{\rm{B}}^H}} \right) \\
 {\rm{s.t.}}~&{\rm{Tr}}\left( {{\bf{t}}_{\rm{W}}{{\bf{W}}_{\rm{b}}}{{\bf{t}}_{\rm{W}}^H}} \right)=0,\label{problem1Bb}\\
 & {\rm{Tr}}\left( {{{\bf{W}}_{\rm{b}}}} \right) \le {P_{{\rm{total}}}},\label{problem1c}\\
 &{{{\bf{W}}_{\rm{b}}}} \underline  \succ  {\bf{0}}.\label{problem1Bd}
 \end{align}
\end{subequations}

Denote ${\bf{W}}_{{\rm{b}}}^*$  as the optimal solutions of \eqref{problem1B}.
Due to relaxation, the rank of  ${\bf{W}}_{{\rm{b}}}^*$  may not equal to one.
Therefore, if ${\rm{rank}}\left( {{\bf{W}}_{{\rm{b}}}^*} \right) = 1$,  the optimal solutions of \eqref{C_20} is ${\bf{W}}_{{\rm{b}}}^*$, and the optimal beamformer  ${{\bf{w}}_{{\rm{b}}}}$   is solved based on the singular value decomposition (SVD), i.e.,${\bf{W}}_{{\rm{b}}}^* = {{\bf{w}}_{{\rm{b}}}}{\bf{w}}_{{\rm{b}}}^H$.
Otherwise,
we propose a  projection approximation
procedure  to obtain a high-quality rank-one solution to \eqref{problem1B}, which is summarized in Algorithm 1.

{\bf {Proposition 1:}} Let ${\bf{W}}_{{\rm{b}}}^*$ denote the optimal solution of
problem \eqref{problem1B}. If ${\rm{rank}}\left( {{\bf{W}}_{{\rm{b}}}^*} \right) > 1$, then the projection approximation
procedure can provide a rank-one solution ${{{\bf{\bar W}}}_{\rm{b}}}$ that satisfies constraint  \eqref{problem1Bb} and  \eqref{problem1c}.

\emph{ Proof:} Please see Appendix A for proof.

\begin{algorithm}[htb]
\caption{  :Projection approximation procedure for problem \eqref{problem1B} }
  \begin{algorithmic}[1]
    \State {{Set}} an SDR solution ${\bf{W}}_{\rm{b}}$, ${\bf{P}}$ denotes the project matrix of vector ${{\bf{W}}_{\rm{b}}^{{1 \mathord{\left/
 {\vphantom {1 2}} \right.
 \kern-\nulldelimiterspace} 2}}{\bf{t}}_{\rm{b}}^H}$, where
    ${\bf{P}} = \frac{{{\bf{W}}_{\rm{b}}^{{1 \mathord{\left/
 {\vphantom {1 2}} \right.
 \kern-\nulldelimiterspace} 2}}{\bf{t}}_{\rm{b}}^H{{\left( {{\bf{W}}_{\rm{b}}^{{1 \mathord{\left/
 {\vphantom {1 2}} \right.
 \kern-\nulldelimiterspace} 2}}{\bf{t}}_{\rm{b}}^H} \right)}^H}}}{{{{\left\| {{\bf{t}}_{\rm{b}}^H{\bf{W}}_{\rm{b}}^{{1 \mathord{\left/
 {\vphantom {1 2}} \right.
 \kern-\nulldelimiterspace} 2}}} \right\|}^2}}}$;
    \State  Construct a new rank one solution ${{{\bf{\bar W}}}_{\rm{b}}} = {\bf{W}}_{\rm{b}}^{{1 \mathord{\left/
 {\vphantom {1 2}} \right.
 \kern-\nulldelimiterspace} 2}}{\bf{PW}}_{\rm{b}}^{{1 \mathord{\left/
 {\vphantom {1 2}} \right.
 \kern-\nulldelimiterspace} 2}}$ that satisfies constraint  \eqref{problem1Bb} and  \eqref{problem1c};
    \State By SVD method, we can obtain ${{{\bf{w}}_{\rm{b}} ^ *}} $ from ${{{\bf{\bar W}}}_{\rm{b}}}$ for problem  \eqref{problem1B}.
  \end{algorithmic}
\end{algorithm}


  \subsubsection{{{Sub-Problem 2. Optimizing ${\bf{q}}$ with Given ${{\mathbf{w}}_{\rm{b}}}$}}}
When we fix ${{\mathbf{w}}_{\rm{b}}}$, problem \eqref{C_14} can be converted to
\begin{align}\label{C_15}
 \mathop {\max }\limits_{{\bf{q}}} ~&{{{{\left| {\left( {{\bf{h}}_{{\rm{IB}}}^H{\bf{Q}}{{\bf{H}}_{{\rm{AI}}}} + {\bf{h}}_{{\rm{AB}}}^H} \right){{\bf{w}}_{\rm{b}}}} \right|}^2}}} \\
 {\rm{s.t.}}~&\eqref{C_14b},\eqref{C_14d}\notag.
 \end{align}

Since
   \begin{subequations}\label{C_16}
\begin{align}
&{\bf{h}}_{{\rm{IB}}}^H{\bf{Q}}{{\bf{H}}_{{\rm{AI}}}} = {{\bf{q}}^H}{\rm{diag}}\left( {{\bf{h}}_{{\rm{IB}}}^H} \right){{\bf{H}}_{{\rm{AI}}}},\\
&{\bf{h}}_{{\rm{IW}}}^H{\bf{Q}}{{\bf{H}}_{{\rm{AI}}}} = {{\bf{q}}^H}{\rm{diag}}\left( {{\bf{h}}_{{\rm{IW}}}^H} \right){{\bf{H}}_{{\rm{AI}}}},
 \end{align}
\end{subequations}
the following equalities hold:
\begin{align}\label{C_17}
{\left| {\left( {{\bf{h}}_{{\rm{IB}}}^H{\bf{Q}}{{\bf{H}}_{{\rm{AI}}}} + {\bf{h}}_{{\rm{AB}}}^H} \right){{\bf{w}}_{\rm{b}}}} \right|^2}={{{\bf{\bar q}}}^H}{{\bf{G}}_{\rm{B}}}{\bf{\bar q}}{\rm{ + }}{h_{\rm{B}}},
 \end{align}
 \begin{align}\label{C_18}
{\left| {\left( {{\bf{h}}_{{\rm{IW}}}^H{\bf{Q}}{{\bf{H}}_{{\rm{AI}}}} + {\bf{h}}_{{\rm{AW}}}^H} \right){{\bf{w}}_{\rm{b}}}} \right|^{\rm{2}}} = {{{\bf{\bar q}}}^H}{{\bf{G}}_{\rm{W}}}{\bf{\bar q}}{\rm{ + }}{h_{\rm{W}}},
 \end{align}
 where ${\bf{\bar q}} = {\left[ {{{\bf{q}}^H},1} \right]^H}$, ${h_{\rm{B}}} \buildrel \Delta \over = {\bf{h}}_{{\rm{AB}}}^H{{\bf{w}}_{\rm{b}}}{\bf{w}}_{\rm{b}}^H{{\bf{h}}_{{\rm{AB}}}}$ and ${h_{\rm{W}}} \buildrel \Delta \over ={\bf{h}}_{{\rm{AW}}}^H{{\bf{w}}_{\rm{b}}}{\bf{w}}_{\rm{b}}^H{{\bf{h}}_{{\rm{AW}}}}$. In addition, ${{\bf{G}}_{\rm{B}}} $ and ${{\bf{G}}_{\rm{W}}} $ are defined in \eqref{GB} and \eqref{GW}, shown at the top of next page, respectively.
 \begin{figure*}[pt]
 \begin{align}\label{GB}
{{\bf{G}}_{\rm{B}}}  \buildrel \Delta \over =  \left[ {\begin{array}{*{20}{c}}
{{\rm{diag}}\left( {{\bf{h}}_{{\rm{IB}}}^H} \right){{\bf{H}}_{{\rm{AI}}}}{{\bf{w}}_{\rm{b}}}{\bf{w}}_{\rm{b}}^H{\bf{H}}_{{\rm{AI}}}^H{\rm{diag}}\left( {{\bf{h}}_{{\rm{IB}}}^H} \right)}&{{\rm{diag}}\left( {{\bf{h}}_{{\rm{IB}}}^H} \right){{\bf{H}}_{{\rm{AI}}}}{{\bf{w}}_{\rm{b}}}{\bf{w}}_{\rm{b}}^H{{\bf{h}}_{{\rm{AB}}}}}\\
{{\bf{h}}_{{\rm{AB}}}^H{{\bf{w}}_{\rm{b}}}{\bf{w}}_{\rm{b}}^H{\bf{H}}_{{\rm{AI}}}^H{\rm{diag}}\left( {{\bf{h}}_{{\rm{IB}}}^H} \right)}&0
\end{array}} \right],
 \end{align}
 \end{figure*}
 \begin{figure*}[pt]
 \begin{align}\label{GW}
{{\bf{G}}_{\rm{W}}} \buildrel \Delta \over = \left[ {\begin{array}{*{20}{c}}
{{\rm{diag}}\left( {{\bf{h}}_{{\rm{IW}}}^H} \right){{\bf{H}}_{{\rm{AI}}}}{{\bf{w}}_{\rm{b}}}{\bf{w}}_{\rm{b}}^H{\bf{H}}_{{\rm{AI}}}^H{\rm{diag}}\left( {{\bf{h}}_{{\rm{IW}}}^H} \right)}&{{\rm{diag}}\left( {{\bf{h}}_{{\rm{IW}}}^H} \right){{\bf{H}}_{{\rm{AI}}}}{{\bf{w}}_{\rm{b}}}{\bf{w}}_{\rm{b}}^H{{\bf{h}}_{{\rm{AW}}}}}\\
{{\bf{h}}_{{\rm{AW}}}^H{{\bf{w}}_{\rm{b}}}{\bf{w}}_{\rm{b}}^H{\bf{H}}_{{\rm{AI}}}^H{\rm{diag}}\left( {{\bf{h}}_{{\rm{IW}}}^H} \right)}&0
\end{array}} \right].
 \end{align}
 \hrulefill
 \end{figure*}

For formula \eqref{C_18},  since ${\bf{W}}_{\rm{b}}$ and ${{\bf{G}}_{\rm{W}}}$ are positive semi-definite matrices,  we simplify the constraint \eqref{C_14b}  to  ${{{\bf{\bar q}}}^H}{{\bf{G}}_{\rm{W}}}{\bf{\bar q}}=0$.
Then, we can rewrite problem \eqref{C_15}  as
    \begin{subequations}\label{problem1}
\begin{align}
 \mathop {\max }\limits_{{{\bf{\bar q}}}} ~&{{{{{\bf{\bar q}}}^H}{{\bf{G}}_{\rm{B}}}{\bf{\bar q}}{\rm{ + }}{h_{\rm{B}}}}}\\
 {\rm{s.t.}}~&{{{\bf{\bar q}}}^H}{{\bf{G}}_{\rm{W}}}{\bf{\bar q}}=0,\\
 &{{{\bf{\bar q}}}^H}{{\bf{E}}_{{m}}}{\bf{\bar q}}=1,\forall m,
 \end{align}
\end{subequations}
where ${{\bf{E}}_{{m}}}$ is an $M+1$ dimensional matrix, and the ${\left[ {{{\bf{E}}_m}} \right]_{i,j}}$ is denoted
as the $\left( {i,j} \right)$th element, satisfies
\begin{align}
{\left[ {{{\bf{E}}_m}} \right]_{i,j}} = \left\{ {\begin{array}{*{20}{c}}
{1,\begin{array}{*{20}{c}}
{}&{}
\end{array}i = j = m,}\\
{0,\begin{array}{*{20}{c}}
{}&{}
\end{array}{\rm {otherwise.}}}
\end{array}} \right.
 \end{align}

 It is still hard to find the optimal solution to  \eqref{problem1}. Next, the SDR is applied
to tackle the non-convexity. Let ${\bf{\bar Q}} = {\bf{\bar q}}{{{\bf{\bar q}}}^H}$, problem \eqref{problem1} is expressed as  its relaxed form without considering
the constraint of ${\rm{rank}}\left( {{\bf{{\bar Q}}}} \right) = 1$, which is given by
    \begin{subequations}\label{problem1A}
\begin{align}
\mathop {\max }\limits_{{\bf{\bar Q}}} ~&  {{\rm{Tr}}\left( {{{\bf{G}}_{\rm{B}}}{\bf{\bar Q}}} \right){\rm{ + }}{h_{\rm{B}}}} \\
{\rm{s}}.{\rm{t}}.&{\rm{Tr}}\left( {{{\bf{G}}_{\rm{W}}}{\bf{\bar Q}}} \right) = 0,\\
&{\rm{Tr}}\left( {{{\bf{E}}_m}{\bf{\bar Q}}} \right) = 1,\forall m,\\
&{\bf{\bar Q}}\underline  \succ  {\bf{0}}.
 \end{align}
\end{subequations}

Problem \eqref{problem1A} is a convex semidefinite programming (SDP) problem, which can be solved based on the interior-point method.
The issue brought by the relaxation of SDR can be similarly handled as we do for problem \eqref{problem1A}.


\subsubsection{{Covert Beamformer Design Algorithm}}

In summary, the optimal covert beamformers of   problem \eqref{C_14}  can be obtained by
  solving   sub-problem $1$ and sub-problem $2$ iteratively, and the overall algorithm  is listed in Algorithm 2.
The complexity of solving sub-problem $1$ and sub-problem $2$ is ${\cal O}\left( {\max {{\left\{ {2,N} \right\}}^4}\sqrt {N} \log \left( {{1 \mathord{\left/
 {\vphantom {1 {{\xi _1}}}} \right.
 \kern-\nulldelimiterspace} {{\xi _1}}}} \right)} \right)$ and ${\cal O}\left( {{{\left( {M+1}  \right)}^{4.5}}\log \left( {{1 \mathord{\left/
 {\vphantom {1 {{\xi _1}}}} \right.
 \kern-\nulldelimiterspace} {{\xi _1}}}}  \right)} \right)$ for each iteration, respectively, where ${\xi _1} > 0$ is the pre-defined accuracy of problem \eqref{C_13} \cite{ZLuo10Semidefinite,Grant09cvx}.
  In Algorithm 2, ${{\bf{1}}_N}$ is denoted as an $N \times 1$ vector with all elements being $1$; $R_{\rm{b}}^{\left( k \right)} = f\left( {{\bf{w}}_{\rm{b}}^{\left( k \right)},{{\bf{q}}^{\left( k \right)}}} \right)$ is denoted as the objective value of \eqref{C_14}, where
${{\bf{w}}_{\rm{b}}^{\left( k \right)}}$ and ${{\bf{q}}^{\left( {k } \right)}}$  are the $k$th iteration variables, and $\epsilon > 0$ is denoted as a threshold.

\begin{algorithm}[htb]
  \caption{: Proposed covert beamformer design algorithm}
  \begin{algorithmic}[1]
    \State {\bf{Initialization:}} Set $k = 0$, ${\bf{w}}_{\rm{b}}^{\left( 0 \right)} = {{\sqrt {{P_{{\rm{total}}}}} {{\bf{h}}_{{\rm{AB}}}}} \mathord{\left/
 {\vphantom {{\sqrt {{P_{{\rm{total}}}}} {{\bf{h}}_{{\rm{AB}}}}} {\left\| {{{\bf{h}}_{{\rm{AB}}}}} \right\|}}} \right.
 \kern-\nulldelimiterspace} {\left\| {{{\bf{h}}_{{\rm{AB}}}}} \right\|}}$, ${{\bf{q}}^{\left( 0 \right)}} = {{\bf{1}}_N}$ and $R_{\rm{b}}^{\left( 0 \right)} = f\left( {{\bf{w}}_{\rm{b}}^{\left( 0 \right)},{{\bf{q}}^{\left( 0 \right)}}} \right)$;
    \State {\bf{repeat}}
     \State Set $k = k + 1$;
    \State  With given ${{\bf{q}}^{\left( {k - 1} \right)}}$, solve problem \eqref{problem1B}  and apply Algorithm $1$ over ${{\bf{W}}_{\rm{b}}^{\left( k \right)}}$ to obtain an approximate solution ${{\bf{w}}_{\rm{b}}^{\left( k \right)}}$;
    \State With given ${{\bf{w}}_{\rm{b}}^{\left( k \right)}}$, solve problem \eqref{problem1A}  and apply
Gaussian randomization over its solution to obtain an approximate solution ${{\bf{q}}^{\left( {k } \right)}}$;
   \State  Set $R_{\rm{b}}^{\left( k \right)} = f\left( {{\bf{w}}_{\rm{b}}^{\left( k \right)},{{\bf{q}}^{\left( k \right)}}} \right)$;
    \State {\bf{until}} $\frac{{R_{\rm{b}}^{\left( k \right)} - R_{\rm{b}}^{\left( {k - 1} \right)}}}{{R_{\rm{b}}^{\left( k \right)}}} < \epsilon  $;
    \State     Output ${{\bf{w}}_{\rm{b}}^{\left( k \right)}}$ and  ${{\bf{q}}^{\left( {k } \right)}}$.
  \end{algorithmic}
\end{algorithm}

 \subsection{Discrete Phase Shifts Design}

For discrete phase shifts,  we
consider that the phase shift at each element of the IRS can only
take a finite number of discrete values, which are equally drawn from
$\left[ {0,2\pi } \right)$. Denote by $L$ the number of bits. Then the set of phase shifts at each element is given by
${\cal {F}} = \left\{ {{\rm{0,}}\Delta \theta , \cdots ,\Delta \theta \left( {K - 1} \right)} \right\}$ where $\Delta \theta  = \frac{{2\pi }}{K}$ and $K = {2^{L}}$.

Similar to question (15), we aim to maximize the achievable covert  rate of Bob  ${R_{\rm{b}}}$ by jointly optimizing the transmit beamforming ${{\bf{w}}_{\rm{b}}}$ and phase shifts ${{\hat \theta }}$,
   subject to the   perfect covert transmission constraint, the total transmit power constraints of Alice and the IRS-related QoS. This is mathematically expressed as
\begin{subequations}\label{C13d}
\begin{align}
\mathop {\max }\limits_{{{\bf{w}}_{\rm{b}}},{{\hat \theta }}} {\rm{ }}&~{R_{\rm{b}}} \hfill \label{C14ad}\\
  {\text{s}}{\text{.t}}{\text{.}}~&D\left( {{p_0}\left\| {{p_1}} \right.} \right)= 0, \hfill \label{C14bd}\\
 & {\left\| {{{\mathbf{w}}_{\rm{b}}}} \right\|^2} \le  {P_{{\rm{total}}}}, \hfill \label{C14cd}\\
 &{\theta _n} \in {\cal {F}}, \forall n,\label{C14d_d}
  \end{align}
\end{subequations}
where ${\hat \theta }  = \left[ {{\theta _1}, \cdots ,{\theta _M}} \right]$.

  Note that problem \eqref{C13d} is   non-convex and  difficult to be optimally solved.
%

Similar to the method of solving  problem \eqref{C_14},   we need to alternately optimize ${{\mathbf{w}}_{\rm{b}}}$ and ${{\hat \theta }}$ for the following two sub-problems.

 \subsubsection{{{Sub-Problem 3. Optimizing ${{\mathbf{w}}_{\rm{b}}}$ with Given ${{\hat \theta }}$}}}

When we fix ${{\hat \theta }}$, problem \eqref{C13d} can be converted to the following problem
\begin{align}\label{C20d}
 \mathop {\max }\limits_{{{\mathbf{w}}_{\rm{b}}}} ~&{\left| {\left( {{\bf{h}}_{{\rm{IB}}}^H{\bf{Q}}{{\bf{H}}_{{\rm{AI}}}} + {\bf{h}}_{{\rm{AB}}}^H} \right){{\bf{w}}_{\rm{b}}}} \right|^2}\\
 {\rm{s.t.}}~&\eqref{C14bd},\eqref{C14cd}\notag.
 \end{align}
 The handling of this problem is the same as that of sub-problem 1.

  \subsubsection{{{Sub-Problem 4. Optimizing ${{\hat \theta }}$ with Given ${{\mathbf{w}}_{\rm{b}}}$}}}

When we fix ${{\mathbf{w}}_{\rm{b}}}$, problem \eqref{C13d} can be converted to
\begin{align}\label{C15d}
 \mathop {\max }\limits_{{{\hat \theta }}} ~&{{{{\left| {\left( {{\bf{h}}_{{\rm{IB}}}^H{\bf{Q}}{{\bf{H}}_{{\rm{AI}}}} + {\bf{h}}_{{\rm{AB}}}^H} \right){{\bf{w}}_{\rm{b}}}} \right|}^2}}} \\
 {\rm{s.t.}}~&\eqref{C14d_d}\notag,
 \end{align}
Since ${{\hat \theta }}$ can only take a fixed value, it is difficult to satisfy the equality requirements of constraint \eqref{C14bd}, so in the sub-problem $4$ we remove this constraint, and optimize ${{\mathbf{w}}_{\rm{b}}}$ through the sub-problem $3$ to ensure  the constraint \eqref{C14bd} is established.

Let $\Phi  = {\rm {diag}}\left( {{\bf{h}}_{{\rm{IB}}}^H} \right){{\bf{H}}_{{\rm{AI}}}}{{\bf{w}}_{\rm{b}}}$, ${\bf{A}} = \Phi {\Phi ^H}$ and ${{{\bf{\bar h}}}_{{\rm{AB}}}} = \Phi {{\bf{w}}_{\rm{b}}^H}{{\bf{h}}_{{\rm{AB}}}}$.
Denote by ${{\bf{A}}_{m,k}}$ and ${{{\bf{\bar h}}}_{{\rm{AB}},m}}$ the $\left( {m,k} \right)$th and $m$th elements in ${\bf{A}}$ and ${{{\bf{\bar h}}}_{{\rm{AB}}}}$, respectively. Then the key to solving \eqref{C15d} by applying alternating optimization lies
in the observation that for a given $m \in \left\{ {1, \cdots ,M} \right\}$, by fixing ${\theta _k}$'s, $\forall k \ne m$, the objective function of \eqref{C15d} is linear with respect to ${e^{j{\theta _m}}}$, which can be written as
\begin{align}\label{C16d}
2{\mathop{\rm Re}\nolimits} \left\{ {{e^{j{\theta _m}}}{\varsigma _m}} \right\} + \sum\limits_{k \ne m}^M {\sum\limits_{i \ne m}^M {{{\bf{A}}_{k,i}}{e^{j\left( {{\theta _k} - {\theta _i}} \right)}}} }  + C,
 \end{align}
where ${\varsigma _m} = \sum\nolimits_{k \ne m}^M {{{\bf{A}}_{m,k}}{e^{ - j{\theta _k}}}}  + {{{\bf{\bar h}}}_{{\rm{AB}},m}} = \left| {{\varsigma _m}} \right|{e^{ - j{\varphi _m}}}$ and $C = {{\bf{A}}_{m,m}} + 2{\mathop{\rm Re}\nolimits} \left\{ {\sum\nolimits_{k \ne m}^M {{e^{j{\theta _k}}}} {{{\bf{\bar h}}}_{{\rm{AB}},k}}} \right\} + {\left\| {{\bf{h}}_{{\rm{AB}}}^H{{\bf{w}}_{\rm{b}}}} \right\|^2}$. Based on \eqref{C16d}, it is not difficult to verify that the optimal $m$th phase shift is given by \cite{WuQ2019Beamforming}
\begin{align}\label{C17d}
\theta _m^ *  = \arg \mathop {\min }\limits_{\theta  \in {\cal { F}}} \left| {\theta  - {\varphi _m}} \right|.
 \end{align}

 \subsubsection{{Discrete Phase Shifts Algorithm}}

In summary, the optimal covert beamformers of   problem \eqref{C13d}  can be obtained by
  solving   sub-problem $3$ and sub-problem $4$ iteratively, and the overall algorithm   is similar to Algorithm $2$.

Note that the algorithm requires a proper choice of initial
discrete phase shifts ${{{\hat \theta }} ^{\left( 0 \right)}}$, which can be obtained by solving the problem (15),
and then quantizing the continuous phase shifts
obtained to the nearest points in ${\cal { F}}$ similarly as \eqref{C17d}.

 \section{Proposed Robust  Covert Transmission for Imperfect WCSI}
In many other cases,  the WCSI    may not be always accessible to Alice because of the potential limited
cooperation between Alice and Willie.
  We consider another practical scenario
where Willie is a regular user with only limited cooperation to Alice. In this case, Alice has imperfect CSI knowledge due to the passive warden \cite{Goeckel16cov} and channel
estimation errors \cite{Shahzad2017Covert,Forouzesh20Communication}.
Here, the imperfect
   WCSI is modeled as \footnote{
   When Willie is completely passive, we may turn to using the Willie's channel distribution information \cite{Zheng20Covert,Forouzesh19arXiv}.}
 \begin{align}
 {{\bf{h}}_{\rm{AW}}} = {{{\bf{\hat h}}}_{\rm{AW}}} + \Delta {{\bf{h}}_{\rm{AW}}},
 \end{align}
 and
  \begin{align}
 {{\bf{h}}_{\rm{IW}}} = {{{\bf{\hat h}}}_{\rm{IW}}} + \Delta {{\bf{h}}_{\rm{IW}}},
 \end{align}
 where ${{{\bf{\hat h}}}_{\rm{AW}}}$  and ${{{\bf{\hat h}}}_{\rm{IW}}}$ denote the   estimated CSI vector between Alice and Willie, between  Willie and IRS, respectively. $\Delta {{\bf{h}}_{\rm{AW}}}$ and $\Delta {{\bf{h}}_{\rm{IW}}}$  denote  the corresponding CSI error
vectors. Moreover,  $\Delta {{\bf{h}}_{\rm{AW}}}$ and $\Delta {{\bf{h}}_{\rm{IW}}}$ are
characterized by an ellipsoidal region, i.e.,
\begin{align}
{{\cal E}_{\rm{AW}}} \buildrel \Delta \over =\left\{ {\Delta {{\bf{h}}_{\rm{AW}}}\left| {\Delta {\bf{h}}_{\rm{AW}}^H{{\bf{C}}_{\rm{AW}}}\Delta {{\bf{h}}_{\rm{AW}}} \le {v_{\rm{AW}}}} \right.} \right\}\label{C_25},
\end{align}
and
\begin{align}
{{\cal E}_{\rm{IW}}} \buildrel \Delta \over =\left\{ {\Delta {{\bf{h}}_{\rm{IW}}}\left| {\Delta {\bf{h}}_{\rm{IW}}^H{{\bf{C}}_{\rm{IW}}}\Delta {{\bf{h}}_{\rm{IW}}} \le {v_{\rm{IW}}}} \right.} \right\}\label{C_26},
\end{align}
where ${{\bf{C}}_{\rm{AW}}} = {{\bf{C}}_{\rm{AW}}^H}\underline  \succ  {\bf{0}}$, ${{\bf{C}}_{\rm{IW}}} = {{\bf{C}}_{\rm{IW}}^H}\underline  \succ  {\bf{0}}$ control the axes of the ellipsoid, and  ${v_{\rm{AW}}}  >  0$, ${v_{\rm{IW}}}  >  0$ determine the volume of the ellipsoid \cite{XYu20Robust,Zhou20Robust}.

\subsection{Willie's Detection Performance}

Based on the model described above, we   investigate  the optimal decision threshold of Willie, and derive the corresponding  false alarm
 and miss detection probabilities. We focus on the worst case scenario for covert transmission, in which the beamformers
${\mathbf{w}}_{\rm{b}}$ is  known by Willie.

 According to the
Neyman-Pearson criterion \cite{Lehmann_2005_Testing}, the likelihood
ratio test
is considered as the the optimal  rule  of  the  detection error  minimization \cite{Lehmann_2005_Testing} , which is expressed as
 \begin{align}\label{criterion}
\frac{{{p_1}\left( {{y_{\rm{w}}}} \right)}}{{{p_0}\left( {{y_{\rm{w}}}} \right)}}\frac{{\mathop  > \limits^{{{\cal D}_1}} }}{{\mathop  < \limits_{{{\cal D}_0}} }}1,
\end{align}
where ${{{\cal D}_1}}$ and ${{{\cal D}_0}}$ are the  binary decisions corresponding to hypotheses ${{{\cal H}_1}}$
and  ${{{\cal H}_0}}$, respectively, as we defined previously.
Furthermore, equation \eqref{criterion} can be reformulated as
\begin{align}\label{phi}
{\left| {{y_{\rm{w}}}} \right|^2} \frac{{\mathop  > \limits^{{{\cal D}_1}} }}{{\mathop  < \limits_{{{\cal D}_0}} }} {\phi ^*}.
\end{align}
In \eqref{phi}, ${\phi ^*}$ is the optimal threshold for ${\left| {{y_{\rm{w}}}} \right|^2}$, which is given by
\begin{align}
{\phi ^*} \buildrel \Delta \over = \frac{{{\lambda _0}{\lambda _1}}}{{{\lambda _1} - {\lambda _0}}}{\rm{ln}}\frac{{{\lambda _1}}}{{{\lambda _0}}}.
\end{align}
Note that, $\lambda _0$ and $\lambda _1$ are given in \eqref{C_8}, and they depend on the beamformer vectors ${\mathbf{w}}_{\rm{b}}$ and the IRS reflect beamforming vector ${\bf{q}}$.

Following \eqref{C_8},   the cumulative density functions (CDFs) of ${\left| {{y_{\rm{w}}}} \right|^2}$ under ${{{\cal H}_{\rm{0}}}}$ and ${{{\cal H}_{\rm{1}}}}$ are respectively given by
\begin{subequations}\label{C_9}
\begin{align}
 &\Pr \left( {{{\left| {{y_{\rm{w}}}} \right|}^2}|{\cal{H}_{\rm{0}}}} \right) = 1 - \exp \left( { - \frac{{{{\left| {{y_{\rm{w}}}} \right|}^2}}}{{{\lambda _0}}}} \right), \\
 &\Pr \left( {{{\left| {{y_{\rm{w}}}} \right|}^2}|{\cal{H}_{\rm{1}}}} \right) = 1 - \exp \left( { - \frac{{{{\left| {{y_{\rm{w}}}} \right|}^2}}}{{{\lambda _1}}}} \right).
 \end{align}
\end{subequations}
Therefore,  based on   the optimal detection threshold ${\phi ^*}$, the false alarm $ P \left( {{{\cal D}_1}\left| {{{\cal H}_0}} \right.} \right)$ and   missed detection probabilities   $P \left( {{{\cal D}_0}\left| {{{\cal H}_1}} \right.} \right)$  are given as
\begin{subequations}
\begin{align}
 &P \left( {{{\cal D}_1}\left| {{{\cal H}_0}} \right.} \right)= {\Pr}\left( {{{\left| {{y_{\rm{w}}}} \right|}^2} \ge {\phi ^*}|{{{\cal H}_{\rm{0}}}}} \right) = {\left( {\frac{{{\lambda _1}}}{{{\lambda _0}}}} \right)^{ - \frac{{{\lambda _1}}}{{{\lambda _1} - {\lambda _0}}}}}, \\
 &P \left( {{{\cal D}_0}\left| {{{\cal H}_1}} \right.} \right) = {\Pr}\left( {{{\left| {{y_{\rm{w}}}} \right|}^2} \le {\phi ^*}|{{{\cal H}_{\rm{1}}}}} \right)  = 1 - {\left( {\frac{{{\lambda _1}}}{{{\lambda _0}}}} \right)^{ - \frac{{{\lambda _0}}}{{{\lambda _1} - {\lambda _0}}}}}.
\end{align}
\end{subequations}

Let use $P \left( {{{\cal D}_1}\left| {{{\cal H}_0}} \right.} \right)$ and $P \left( {{{\cal D}_0}\left| {{{\cal H}_1}} \right.} \right)$ to characterize the ideal detection performance of Willie. Such results can be used as
the theoretical benchmark to measure the covert performance
of the robust beamforming design. We will further discuss the detection performance of Willie in the next section.

 In practice, it is common that the obtained CSI is corrupted by certain estimation errors \cite{Wornell16,Bloch16}, which makes the perfect covert transmission, i.e., $D\left( {{p_0}\left\| {{p_1}} \right.} \right) = 0$,  difficult to be achieved. Thus, we apply adopting     $D\left( {{p_0}\left\| {{p_1}} \right.} \right) \le 2{\varepsilon ^2}$ and
$D\left( {{p_1}\left\| {{p_0}} \right.} \right) \le 2{\varepsilon ^2}$ given by \eqref{Dp0p1} as covertness constraints\cite{Bash13,Wornell16,Bloch16,Yan2019Gaussian}.

\subsection{ Case of $D\left( {{p_0}\left\| {{p_1}} \right.} \right) \le 2{\varepsilon ^2}$}


To be specific, we aim to maximize ${R_{\rm{b}}}$ via the joint design of the beamformers  ${{\bf{w}}_{\rm{b}}}$ and the IRS reflect beamforming vector ${\bf{q}}$, under the IRS-related QoS, the covertness constraint and the total power constraint.
 Mathematically,  the robust covert rate maximization problem  is formulated as
\begin{subequations}\label{problem2}
\begin{align}
\mathop {\max }\limits_{{{\bf{w}}_{\rm{b}}},{\bf{q}}} {\rm{ }}&~{R_{\rm{b}}} \hfill \\
  {\text{s}}{\text{.t}}{\text{.}}~&D\left( {{p_0}\left\| {{p_1}} \right.} \right) \le 2{\varepsilon ^2}, \hfill \label{problem2a}\\
 & {\left\| {{{\mathbf{w}}_{\rm{b}}}} \right\|^2} \le  {P_{{\rm{total}}}}, \hfill \label{problem2b}\\
 &\left| {{q_m}} \right| = 1,\forall m,\label{problem2c}\\
&{{\bf{h}}_{\rm{IW}}} = {{{\bf{\hat h}}}_{\rm{IW}}} + \Delta {{\bf{h}}_{\rm{IW}}},\label{problem2d}\\
 &{{\bf{h}}_{\rm{AW}}} = {{{\bf{\hat h}}}_{\rm{AW}}} + \Delta {{\bf{h}}_{\rm{AW}}}.\label{problem2e}
\end{align}
\end{subequations}

It can be seen that  problem \eqref{problem2} is nonconvex, which is difficult
to solve it directly.
To tackle
this issue, we reformulate the  covertness constraint \eqref{problem2a} by exploiting the property  of the function $f\left( x \right) = \ln x + \frac{1}{x} - 1$ for $x > 0$.
Specifically,    the covertness constraint $D\left( {{p_0}\left\| {{p_1}} \right.} \right) = \ln \frac{{{\lambda _1}}}{{{\lambda _0}}} + \frac{{{\lambda _0}}}{{{\lambda _1}}} - 1 \le 2{\varepsilon ^2}$ can be  equivalently transformed as
 \begin{align}
\bar a \le \frac{{{\lambda _1}}}{{{\lambda _0}}} \le \bar b\label{barab},
 \end{align}
where    $\bar a$ and $\bar b$ are the two roots of the equation $\ln \frac{{{\lambda _1}}}{{{\lambda _0}}} + \frac{{{\lambda _0}}}{{{\lambda _1}}} - 1 = 2{\varepsilon ^2}$.
Therefore, constraint \eqref{problem2a} can   be equivalently reformulated as
\begin{align}
&\bar a \le \frac{{{\left| {\left( {{\bf{h}}_{{\rm{IW}}}^H{\bf{Q}}{{\bf{H}}_{{\rm{AI}}}} + {\bf{h}}_{{\rm{AW}}}^H} \right){{\bf{w}}_{\rm{b}}}} \right|^{\rm{2}}} + \sigma _{\rm{w}}^2}}{{\sigma _{\rm{w}}^2}} \le \bar b.\label{C_39}
 \end{align}

Here, due to $\Delta {{\bf{h}}_{\rm{AW}}} \in {{\cal{E}}_{\rm{AW}}}$ and $\Delta {{\bf{h}}_{\rm{IW}}} \in {{\cal{E}}_{\rm{IW}}}$, there are infinite choices for
$\Delta {{\bf{h}}_{\rm{AW}}}$ or $\Delta {{\bf{h}}_{\rm{IW}}}$, in constraint \eqref{problem2d} and \eqref{problem2e}, respectively. This makes the problem \eqref{problem2} non-convex and difficult to be solved.
To overcome this challenge, a relaxation and restriction method are proposed. Specifically,
in the relaxation step, the nonconvex
problem is converted into a convex
SDP;  in
the restriction step, a finite number of linear matrix
inequalities (LMIs) are used to reformulate infinite number of complex
constraints.

Similar to the idea of solving  problem \eqref{C_14},  firstly, we can alternately optimize ${{\mathbf{w}}_{\rm{b}}}$ and ${\bf{q}}$ for problem \eqref{problem2}.

\subsubsection{{{Sub-Problem 5. Optimizing ${{\mathbf{w}}_{\rm{b}}}$ with given ${\bf{q}}$}}}

We first optimize the beamformer ${{\mathbf{w}}_{\rm{b}}}$  by fixing ${\bf{q}}$
 under constraints \eqref{problem2a}, \eqref{problem2b}, \eqref{problem2d} and \eqref{problem2e}.
For mathematical convenience, we define ${{\bf{g}}_{\rm{B}}}{\rm{ = }}{\left[ {\begin{array}{*{20}{c}}
{{\bf{h}}_{{\rm{IB}}}^H{\bf{Q}}{{\bf{H}}_{{\rm{AI}}}}}&{{\bf{h}}_{{\rm{AB}}}^H}
\end{array}} \right]^H}$,
${{\bf{g}}_{\rm{W}}}{\rm{ = }}{\left[ {\begin{array}{*{20}{c}}
{{\bf{h}}_{{\rm{IW}}}^H{\bf{Q}}{{\bf{H}}_{{\rm{AI}}}}}&{{\bf{h}}_{{\rm{AW}}}^H}
\end{array}} \right]^H}$
and ${{{\bf{\hat w}}}_{\rm{b}}}{\rm{ = }}\left[ {\begin{array}{*{20}{c}}
{{{\bf{w}}_{\rm{b}}}}\\
{{{\bf{w}}_{\rm{b}}}}
\end{array}} \right]$.
Then, the problem can be given as
\begin{subequations}\label{problem2sub1}
\begin{align}
\mathop {\max }\limits_{{{{\bf{\hat w}}}_{\rm{b}}}} {\rm{ }}~&{{{{\left| {{\bf{g}}_{\rm{B}}^H{{{\bf{\hat w}}}_{\rm{b}}}} \right|}^2}}} \\
  {\text{s}}{\text{.t}}{\text{.}}~&\sigma _{\rm{w}}^2\left( {\bar a{\rm{ - 1}}} \right) \le {\left| {{\bf{g}}_{\rm{W}}^H{{{\bf{\hat w}}}_{\rm{b}}}} \right|^{\rm{2}}} \le \sigma _{\rm{w}}^2\left( {\bar b{\rm{ - 1}}} \right), \label{constraint1}\hfill \\
&{\left\| {{{\bf{e}}_1}{{{\bf{\hat w}}}_{\rm{b}}}} \right\|^2} \le {P_{{\rm{total}}}},\\
&{{\bf{g}}_{\rm{W}}}  = {{{\bf{\hat g}}}_{\rm{W}}} + \Delta {{\bf{g}}_{\rm{W}}},\Delta {{\bf{g}}_{\rm{W}}} \in {{\cal E}_{\rm{W}}},
\end{align}
\end{subequations}
where  ${{\bf{e}}_1} = \left[ {\underbrace {1, \cdots ,{\rm{1}}}_N{\rm{,}}\underbrace {{\rm{0,}} \cdots {\rm{,0}}}_N} \right]$, ${{\cal E}_{\rm{W}}} \buildrel \Delta \over =\left\{ {\Delta {{\bf{g}}_{\rm{W}}}\left| {\Delta {\bf{g}}_{\rm{W}}^H{{\bf{C}}_{\rm{W}}}\Delta {{\bf{g}}_{\rm{W}}} \le {v_{\rm{W}}}} \right.} \right\}$, and  ${v_{\rm{W}}} = {v_{{\rm{AW}}}} + {v_{{\rm{IW}}}}$.

To handle the non-convexity issue, we  relax  the constraint \eqref{constraint1} to a convex form by applying SDR as well. By relaxing ${\widehat {\bf{W}}_{\rm{b}}} = {{{\bf{\hat w}}}_{\rm{b}}}{\bf{\hat w}}_{\rm{b}}^H$
to ${\widehat {\bf{W}}_{\rm{b}}} \succeq {\bf{0}}$, the constraint can be  equivalently re-expressed as
\begin{subequations}
\begin{align}
&\Delta {\bf{g}}_{\rm{W}}^H{\widehat {\bf{W}}_{\rm{b}}}\Delta {{\bf{g}}_{\rm{W}}} + 2\Delta {\bf{g}}_{\rm{W}}^H{\widehat {\bf{W}}_{\rm{b}}}{{{\bf{\hat g}}}_{\rm{W}}} + {\bf{\hat g}}_{\rm{W}}^H{\widehat {\bf{W}}_{\rm{b}}}{{{\bf{\hat g}}}_{\rm{W}}} \ge \sigma _{\rm{w}}^2\left( {\bar a - 1} \right)\label{W1a},\\
& \Delta {\bf{g}}_{\rm{W}}^H{\widehat {\bf{W}}_{\rm{b}}}\Delta {{\bf{g}}_{\rm{W}}} + 2\Delta {\bf{g}}_{\rm{W}}^H{\widehat {\bf{W}}_{\rm{b}}}{{{\bf{\hat g}}}_{\rm{W}}} + {\bf{\hat g}}_{\rm{W}}^H{\widehat {\bf{W}}_{\rm{b}}}{{{\bf{\hat g}}}_{\rm{W}}}\le \sigma _{\rm{w}}^2\left( {\bar b - 1} \right)\label{W1b},
 \end{align}
 \end{subequations}
 where  ${{{\bf{\hat g}}}_{\rm{W}}} \buildrel \Delta \over ={\left[ {\begin{array}{*{20}{c}}
{{\bf{\hat h}}_{{\rm{IW}}}^H{\bf{Q}}{{\bf{H}}_{{\rm{AI}}}}}&{{\bf{\hat h}}_{{\rm{AW}}}^H}
\end{array}} \right]^H}$  and
$\Delta {{\bf{g}}_{\rm{W}}} \buildrel \Delta \over = {\left[ {\begin{array}{*{20}{c}}
{\Delta {\bf{h}}_{{\rm{IW}}}^H{\bf{Q}}{{\bf{H}}_{{\rm{AI}}}}}&{\Delta {\bf{h}}_{{\rm{AW}}}^H}
\end{array}} \right]^H}$.

By applying SDR, we  ignore the  rank-one constraints of   ${\widehat {\bf{W}}_{\rm{b}}}$,
   which is similar to the approach used in \eqref{C_20} and  \eqref{problem1B}. Then, problem \eqref{problem2sub1}  can be relaxed as follows
\begin{subequations}\label{2sub1sdr}
\begin{align}
\mathop {\max }\limits_{{\widehat {\bf{W}}_{\rm{b}}}} {\rm{ }}~&{\rm{Tr}}\left( {{\bf{g}}_{\rm{B}}^H{{\widehat {\bf{W}}}_{\rm{b}}}{\bf{g}}_{\rm{B}}} \right)\\
  {\text{s}}{\text{.t}}{\text{.}}~&{\rm{Tr}}\left( {{{\bf{e}}_1}{{\widehat {\bf{W}}}_{\rm{b}}}{\bf{e}}_1^H} \right)\le  {P_{{\rm{total}}}},\label{2sub1sdra}\\
 &{\widehat {\bf{W}}_{\rm{b}}} \underline  \succ  {\bf{0}},\label{2sub1sdrc}\\
  &\Delta {{\bf{g}}_{\rm{W}}} \in {{\cal E}_{\rm{W}}},\label{2sub1sdrb}\\
&\eqref{W1a}, \eqref{W1b}\notag.
 \end{align}
\end{subequations}

It is worth pointing out that the SDR problem \eqref{2sub1sdr} is quasi-convex as the objective
function and constraints are linear in  ${\widehat {\bf{W}}_{\rm{b}}}$. However, problem \eqref{2sub1sdr}
is still computationally intractable because it involves an infinite
number of constraints due to $\Delta {{\bf{g}}_{\rm{W}}} \in {{\cal E}_{\rm{W}}}$.

 In the following, we employ S-Lemma to recast the infinitely many constraints
  as a certain set of LMIs, which  is a tractable  approximation.

\textbf{Lemma 1 (S-Lemma\cite{DWKRobust14})}: Let ${{\bf{A}}_m} \in {{\mathbb{H}}^N}$, ${{\bf{b}}_m} \in {{\mathbb{C}}^{N \times 1}}$, and ${c_m} \in {{\mathbb{R}}^{1 \times 1}}$. Denote
a function ${f_m}\left( x \right),m \in \left\{ {1,2} \right\},x \in {{\mathbb{C}}^{N \times 1}}$, we have
\begin{align}
{f_m}\left( x \right) = {{\bf{x}}^H}{{\bf{A}}_m}{\bf{x}} +
2{\rm Re}\left\{ {{\bf{b}}_m^{\rm{H}}{\bf{x}}} \right\} + {c_m}.
\end{align}
Then, ${f_1}\left( x \right) \le 0 \Rightarrow {f_2}\left( x \right) \le 0$ holds if and only if there
exits a variable ${\eta }  \ge 0$ such that
\begin{align}
{\eta } \left[ {\begin{array}{*{20}{c}}
{{{\bf{A}}_1}}&{{{\bf{b}}_1}}\\
{{\bf{b}}_1^{{H}}}&{{c_1}}
\end{array}} \right] - \left[ {\begin{array}{*{20}{c}}
{{{\bf{A}}_2}}&{{{\bf{b}}_2}}\\
{{\bf{b}}_2^{{H}}}&{{c_2}}
\end{array}} \right]\underline  \succ  {\bf{0}}\label{spro}.
\end{align}

Consequently, by using S-Lemma,  constraints \eqref{W1a} and \eqref{W1b} can be respectively given  as a finite number of LMIs:
\begin{small}
\begin{subequations}
\begin{align}
\left[ {\begin{array}{*{20}{c}}
{{{\widehat {\bf{W}}}_{\rm{b}}} + {\eta _1}{{\bf{C}}_{\rm{w}}}}&{{{\widehat {\bf{W}}}_{\rm{b}}}{{{\bf{\hat g}}}_{\rm{W}}}}\\
{{\bf{\hat g}}_{\rm{W}}^H{{\widehat {\bf{W}}}_{\rm{b}}}}&{{\bf{\hat g}}_{\rm{W}}^H{{\widehat {\bf{W}}}_{\rm{b}}}{{{\bf{\hat g}}}_{\rm{W}}} - \sigma _{\rm{w}}^2\left( {\bar a - 1} \right) - {\eta _1}{v_{\rm{W}}}}
\end{array}} \right]\underline  \succ  {\bf{0}}\label{slemma1},\\
\left[ {\begin{array}{*{20}{c}}
{ - {{\widehat {\bf{W}}}_{\rm{b}}} + {\eta _2}{{\bf{C}}_{\rm{w}}}}&{ - {{\widehat {\bf{W}}}_{\rm{b}}}{{{\bf{\hat g}}}_{\rm{W}}}}\\
{ - {\bf{\hat g}}_{\rm{W}}^H{{\widehat {\bf{W}}}_{\rm{b}}}}&{ - {\bf{\hat g}}_{\rm{W}}^H{{\widehat {\bf{W}}}_{\rm{b}}}{{{\bf{\hat g}}}_{\rm{W}}} + \sigma _{\rm{w}}^2\left( {\bar b - 1} \right) - {\eta _2}{v_{\rm{W}}}}
\end{array}} \right]\underline  \succ  {\bf{0}}\label{slemma2}.
\end{align}
\end{subequations}
\end{small}

Therefore, we obtain the conservative approximation of \eqref{2sub1sdr} as follows:
\begin{align}\label{2sub1sdr2}
\mathop {\max }\limits_{{\widehat {\bf{W}}_{\rm{b}}}} {\rm{ }}~&{\rm{Tr}}\left( {{{\bf{g}}_{\rm{B}}^H}{{\widehat {\bf{W}}}_{\rm{b}}}{\bf{g}}_{\rm{B}}} \right)\\
{\rm{s}}.{\rm{t}}.~&\eqref{2sub1sdra}, \eqref{2sub1sdrc}, \eqref{slemma1}, \eqref{slemma2}\notag.
 \end{align}

Problem \eqref{2sub1sdr2} is a convex SDP problem which can be optimally solved with the interior-point method.
Similarly,   let ${\widehat {\bf{W}}_{\rm{b}}}^{opt}$  denotes the optimal solutions of problem \eqref{2sub1sdr2}.
If ${\rm{rank}}\left( {{\widehat {\bf{W}}_{\rm{b}}}^{opt}} \right) = 1$, ${\widehat {\bf{W}}_{\rm{b}}}^{opt}$ is the optimal solutions of problem \eqref{2sub1sdr2}, and the optimal beamformer  ${{{\bf{\hat w}}}_{\rm{b}}}$  can be obtained by SVD, i.e.,${\widehat {\bf{W}}_{\rm{b}}}^{opt} = {{{\bf{\hat w}}}_{\rm{b}}}{{{\bf{\hat w}}}_{\rm{b}}}^H$.
Otherwise, if ${\rm{rank}}\left( {{\widehat {\bf{W}}_{\rm{b}}}^{opt}} \right) > 1$,
   the   Gaussian randomization procedure \cite{ZLuo10Semidefinite}   is adopted to produce a high-quality rank-one solution to \eqref{2sub1sdr2}.

  \subsubsection{{{Sub-Problem 6. Optimizing ${\bf{q}}$ with given ${{\mathbf{w}}_{\rm{b}}}$}}}

Now we consider the design of ${\bf{q}}$ on the basis of fixing  ${{\mathbf{w}}_{\rm{b}}}$.
In this case, the problem \eqref{problem2} can be converted into the following form:
 \begin{subequations}\label{problem2sub2}
\begin{align}
 \mathop {\max }\limits_{{\bf{q}}} ~&{{{{\left| {\left( {{\bf{h}}_{{\rm{IB}}}^H{\bf{Q}}{{\bf{H}}_{{\rm{AI}}}} + {\bf{h}}_{{\rm{AB}}}^H} \right){{\bf{w}}_{\rm{b}}}} \right|}^2}}}\\
{\rm{s.t.}}~&\bar a \le \frac{{{\left| {\left( {{\bf{h}}_{{\rm{IW}}}^H{\bf{Q}}{{\bf{H}}_{{\rm{AI}}}} + {\bf{h}}_{{\rm{AW}}}^H} \right){{\bf{w}}_{\rm{b}}}} \right|^{\rm{2}}} + \sigma _{\rm{w}}^2}}{{\sigma _{\rm{w}}^2}} \le \bar b,\label{problem2sub2a}\\
 &\left| {{q_m}} \right| = 1,\forall m.\label{problem2sub2b}
 \end{align}
\end{subequations}
Since \eqref{problem2d} and \eqref{problem2e} have been discussed in sub-problem 5, we do not consider these two constraints in sub-problem 6. As a result, the processing method is the same as that of question \eqref{C_15}.
The SDR is applied
to tackle the non-convexity with ${\bf{\bar Q}} = {\bf{\bar q}}{{{\bf{\bar q}}}^H}$ and ${\bf{\bar q}} = {\left[ {{{\bf{q}}^H},1} \right]^H}$. Then, \eqref{problem2sub2} is expressed as its relaxed form without considering
the ${\rm{rank}}\left( {{\bf{{\bar Q}}}} \right) = 1$ constraint, which is given by
    \begin{subequations}\label{2sub2sdr}
\begin{align}
\mathop {\max }\limits_{{\bf{\bar Q}}} ~& {{\rm{Tr}}\left( {{{\bf{G}}_{\rm{B}}}{\bf{\bar Q}}} \right){\rm{ + }}{h_{\rm{B}}}} \\
{\rm{s}}.{\rm{t}}.&\sigma _{\rm{w}}^2\left( {\bar a - 1} \right) \le {\rm{Tr}}\left( {{{\bf{G}}_{\rm{W}}}{\bf{\bar Q}}} \right){\rm{ + }}{h_{\rm{W}}}\le \sigma _{\rm{w}}^2\left( {\bar b - 1} \right),\\
&{\rm{Tr}}\left( {{{\bf{E}}_m}{\bf{\bar Q}}} \right) = 1,\forall m,\\
&{\bf{\bar Q}}\underline  \succ  {\bf{0}}.
 \end{align}
\end{subequations}

Problem \eqref{2sub2sdr} is a convex SDP problem, which  can be optimally solved with the interior-point method.
Also, we may apply the similar technique as we do for problem  \eqref{2sub1sdr2} to deal with the issue brought by the relaxation of SDR.

\subsubsection{ {Robust covert beamformers design algorithm}}

 In short, the robust covert beamformers in problem \eqref{problem2} can be design by solving sub-problem 5 and sub-problem 6  alternately, and the  overall  algorithm  is presented in
Algorithm 3.
Similar to problem \eqref{C_13} , the complexity of sub-problem $5$
is ${\cal O}\left( {\max {{\left\{ {3,N} \right\}}^4}\sqrt {N} \log \left( {{1 \mathord{\left/
 {\vphantom {1 {{\xi _1}}}} \right.
 \kern-\nulldelimiterspace} {{\xi _2}}}} \right)} \right)$ for each iteration, and the complexity of sub-problem $6$ is ${\cal O}\left( {{{\left( {M+1}  \right)}^{4.5}}\log \left( {{1 \mathord{\left/
 {\vphantom {1 {{\xi _1}}}} \right.
 \kern-\nulldelimiterspace} {{\xi _2}}}}  \right)} \right)$
for each iteration. ${\xi _2} > 0$ is the pre-defined accuracy of problem \eqref{problem2} \cite{ZLuo10Semidefinite,Grant09cvx}.
Here, 
$R_{\rm{b}}^{\left( k \right)} = f\left( {{\bf{w}}_{\rm{b}}^{\left( k \right)},{{\bf{q}}^{\left( k \right)}}} \right)$ is denoted as  the objective value of  \eqref{problem2}, where ${{\bf{w}}_{\rm{b}}^{\left( k \right)}}$ and ${{\bf{q}}^{\left( {k } \right)}}$ are the $k$th iteration variables.

\begin{algorithm}[htb]
  \caption{: Proposed   robust covert beamformers design algorithm}
  \label{alg2:Alternate iteration}
  \begin{algorithmic}[1]
    \State {\bf{Initialization:}} Set $k = 0$, ${\bf{w}}_{\rm{b}}^{\left( 0 \right)} = \frac{{\sqrt {{P_{{\rm{total}}}}} {{\bf{h}}_{{\rm{AB}}}}}}{{\left\| {{{\bf{h}}_{{\rm{AB}}}}} \right\|}}$, ${{\bf{q}}^{\left( 0 \right)}} = {{\bf{1}}_N}$, and $R_{\rm{b}}^{\left( 0 \right)} = f\left( {{\bf{w}}_{\rm{b}}^{\left( 0 \right)},{{\bf{q}}^{\left( 0 \right)}}} \right)$;
    \State {\bf{repeat}}
     \State Set $k = k + 1$;
    \State  With given ${{\bf{q}}^{\left( {k - 1} \right)}}$, solve problem \eqref{2sub1sdr2} and apply Gaussian randomization over its solution to obtain an approximate solution ${{\bf{w}}_{\rm{b}}^{\left( k \right)}}$;
    \State With given ${{\bf{w}}_{\rm{b}}^{\left( k \right)}}$, solve problem \eqref{2sub2sdr} and apply Gaussian randomization over ${\bf{\bar Q}}^{\left( k \right)}$ to obtain an approximate solution ${{\bf{q}}^{\left( {k } \right)}}$;
   \State  Set $R_{\rm{b}}^{\left( k \right)} = f\left( {{\bf{w}}_{\rm{b}}^{\left( k \right)},{{\bf{q}}^{\left( k \right)}}} \right)$;
    \State {\bf{until}} $\frac{{R_{\rm{b}}^{\left( k \right)} - R_{\rm{b}}^{\left( {k - 1} \right)}}}{{R_{\rm{b}}^{\left( k \right)}}} < \epsilon  $;
        \State   Output ${{\bf{w}}_{\rm{b}}^{\left( k \right)}}$ and  ${{\bf{q}}^{\left( {k } \right)}}$.
  \end{algorithmic}
\end{algorithm}

\subsection{Case of $D\left( {{p_1}\left\| {{p_0}} \right.} \right) \le 2{\varepsilon ^2}$}
In this subsection, the case with constraint $D\left( {{p_1}\left\| {{p_0}} \right.} \right) \le 2{\varepsilon ^2}$ is considered. The   corresponding robust covert rate maximization problem is  given by
\begin{subequations}\label{problem3}
\begin{align}
\mathop {\max }\limits_{{{\bf{w}}_{\rm{b}}},{\bf{q}}} {\rm{ }}&{R_{\rm{b}}}\left( {{{\bf{w}}_{{\rm{c}},1}}},{{{\bf{w}}_{{\rm{b}}}}} \right) \hfill \\
  {\text{s}}{\text{.t}}{\text{.}}~&D\left( {{p_1}\left\| {{p_0}} \right.} \right) \le 2{\varepsilon ^2}, \hfill \label{problem3a}\\
 & {\left\| {{{\mathbf{w}}_{\rm{b}}}} \right\|^2} \le  {P_{{\rm{total}}}}, \hfill \label{problem3b}\\
 &\left| {{q_m}} \right| = 1,\forall m,\label{problem3c}\\
&{{\bf{h}}_{\rm{IW}}} = {{{\bf{\hat h}}}_{\rm{IW}}} + \Delta {{\bf{h}}_{\rm{IW}}},\label{problem3d}\\
 &{{\bf{h}}_{\rm{AW}}} = {{{\bf{\hat h}}}_{\rm{AW}}} + \Delta {{\bf{h}}_{\rm{AW}}},\label{problem3e}
\end{align}
\end{subequations}
where $D\left( {{p_1}\left\| {{p_0}} \right.} \right) = \ln \left( {{{{\lambda _0}} \mathord{\left/
 {\vphantom {{{\lambda _0}} {{\lambda _1}}}} \right.
 \kern-\nulldelimiterspace} {{\lambda _1}}}} \right) + {{{\lambda _{\rm{1}}}} \mathord{\left/
 {\vphantom {{{\lambda _{\rm{1}}}} {{\lambda _{\rm{0}}}}}} \right.
 \kern-\nulldelimiterspace} {{\lambda _{\rm{0}}}}} - 1$.

Note that problem \eqref{problem3}
 is similar to problem \eqref{problem2} except for the covertness constraint.
The covertness constraint $D\left( {{p_1}\left\| {{p_0}} \right.} \right) = \ln \left( {{{{\lambda _0}} \mathord{\left/
 {\vphantom {{{\lambda _0}} {{\lambda _1}}}} \right.
 \kern-\nulldelimiterspace} {{\lambda _1}}}} \right) + {{{\lambda _{\rm{1}}}} \mathord{\left/
 {\vphantom {{{\lambda _{\rm{1}}}} {{\lambda _{\rm{0}}}}}} \right.
 \kern-\nulldelimiterspace} {{\lambda _{\rm{0}}}}} - 1 \le 2{\varepsilon ^2}$ can be  equivalently transformed as
\begin{align}
\bar c \le \frac{{{\lambda _0}}}{{{\lambda _1}}} \le \bar d,
 \end{align}
 where $\bar c= {\bar a}$ and $\bar d= {\bar b}$, are the two roots of the equation $\ln \left( {{{{\lambda _0}} \mathord{\left/
 {\vphantom {{{\lambda _0}} {{\lambda _1}}}} \right.
 \kern-\nulldelimiterspace} {{\lambda _1}}}} \right) + {{{\lambda _{\rm{1}}}} \mathord{\left/
 {\vphantom {{{\lambda _{\rm{1}}}} {{\lambda _{\rm{0}}}}}} \right.
 \kern-\nulldelimiterspace} {{\lambda _{\rm{0}}}}} - 1 = 2{\varepsilon ^2}$.

Similar to the previous subsection, we  apply the alternate iteration, relaxation and restriction approach to solve problem \eqref{problem3}. We omit the detailed derivations for brevity. Although the methods are similar, the achievable covert rates  are quite different under the two covertness constraints. We will illustrate and discuss this issue in   the next section.
\section{Numerical Results}
The numerical results  about the performance of the proposed
covert beamformers design and  robust beamformers design methods for covert communications are presented and discussed.
In our simulations, we set  the number of   antennas  at  Alice to $4$, i.e., $N=4$, and assume that  $M = 4$. The noise
variances of Bob and Willie are $\sigma _{\rm{b}}^2=\sigma _{\rm{w}}^2=-80\rm{dBm}$.
Alice, Bob, Willie,
and the IRS are located at $\left( {0,3} \right)$, $\left( {8,0} \right)$, $\left( {5,0} \right)$, and
$\left( {10,3} \right)$ in meter $\left( {\rm{m}} \right)$ in a two-dimensional area, respectively \cite{Hong20Artificial,Hossain20Intelligent}.

In our simulations, we set ${\zeta _0} =  - 30{\rm{dB}}$.
The path loss exponents of the Alice-Willie link, the Alice-Bob link, the IRS-Willie link, and the IRS-Bob link are ${\alpha _{{\rm{AW}}}} ={\alpha _{{\rm{AB}}}} ={\alpha _{{\rm{IW}}}} = {\alpha _{{\rm{IB}}}} = 3$. For the Alice-IRS link, the
path-loss exponent is  ${\alpha _{{\rm{AI}}}} = 2.2$, which means that the IRS is well-located, and the path loss is negligible in this link.

\subsection{Evaluation for Scenario 1}
Let's evaluate the proposed methods in scenario 1, namely, Alice with perfect WCSI.
First of all,  the numerical results are presented to compare the performance of the proposed covert beamformer design, discrete phase shifts design and the case without IRS beamformer design, which means that no IRS is involved in the system (let ${\bf{q}}={\bf{0}}$ and only design ${{\mathbf{w}}_{\rm{b}}}$ according to problem  \eqref{C_20}).


 \begin{figure}[h]
          \centering
      \includegraphics[width=8cm]{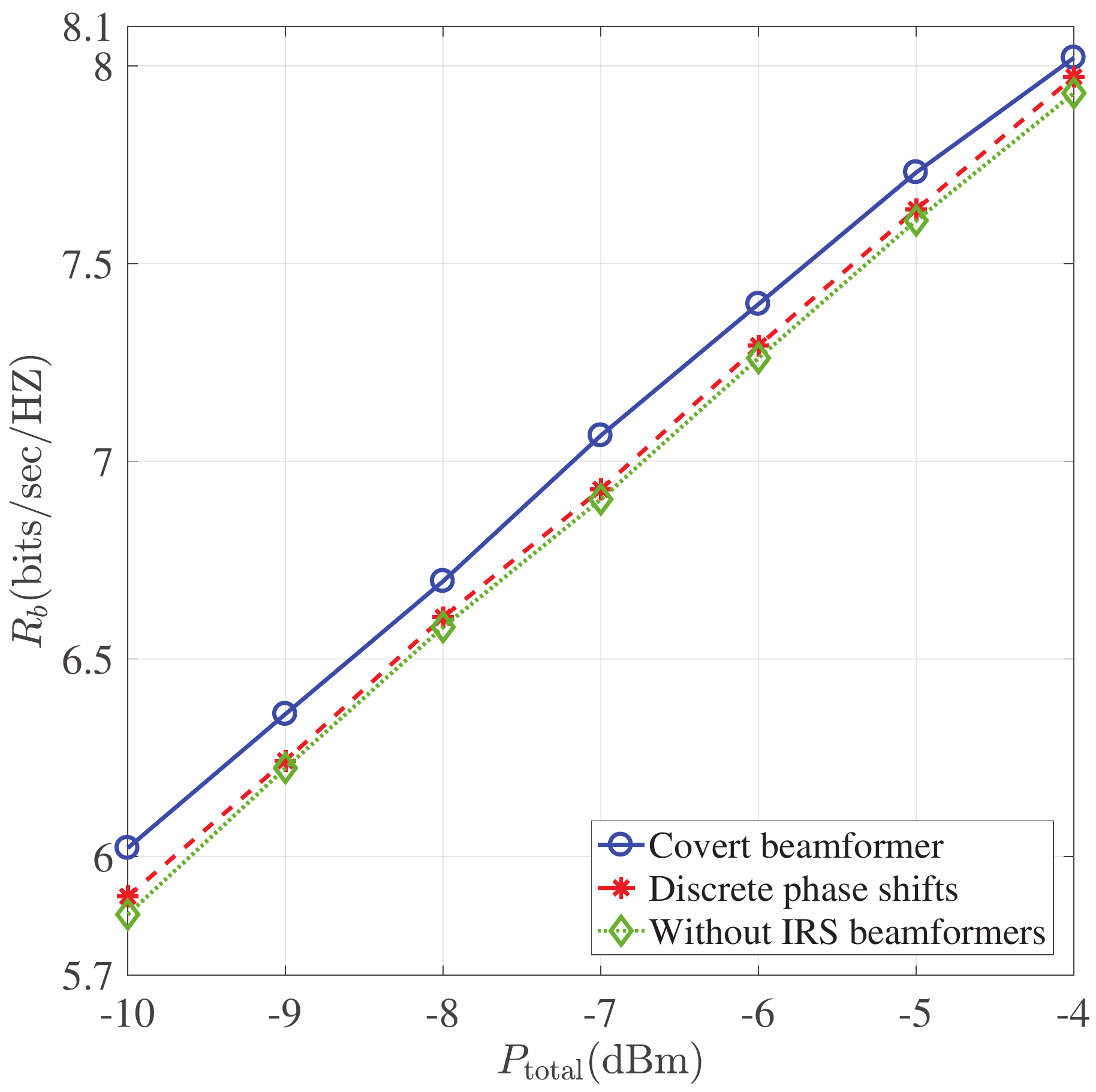}
 \caption{~$R_{\rm{b}}$ (bits/sec/Hz)  versus $P_{\rm{total}}$ (dBW).}
    \label{Ptotal_rb} 
\end{figure}

Fig. \ref{Ptotal_rb}  investigates the covert rate of Bob $R_{\rm{b}}$   versus the total transmit power $P_{\rm {total}}$, where $K$=4.
  We can see that  the covert rate of Bob $R_{\rm{b}}$ increases as the transmit power of Alice $P_{\rm {total}}$ increases.
 More specific,  $R_{\rm{b}}$  of the
 covert beamforming design is the highest among the three design methods, while $R_{\rm{b}}$  of the without IRS design is the lowest.
  This is because that the reflected signal by the IRS and the direct signal can be better added constructively at Bob while destructively at Willie after continuously optimizing the IRS-related phase shifts.

\begin{figure}[h]
      \centering
	\includegraphics[width=8cm]{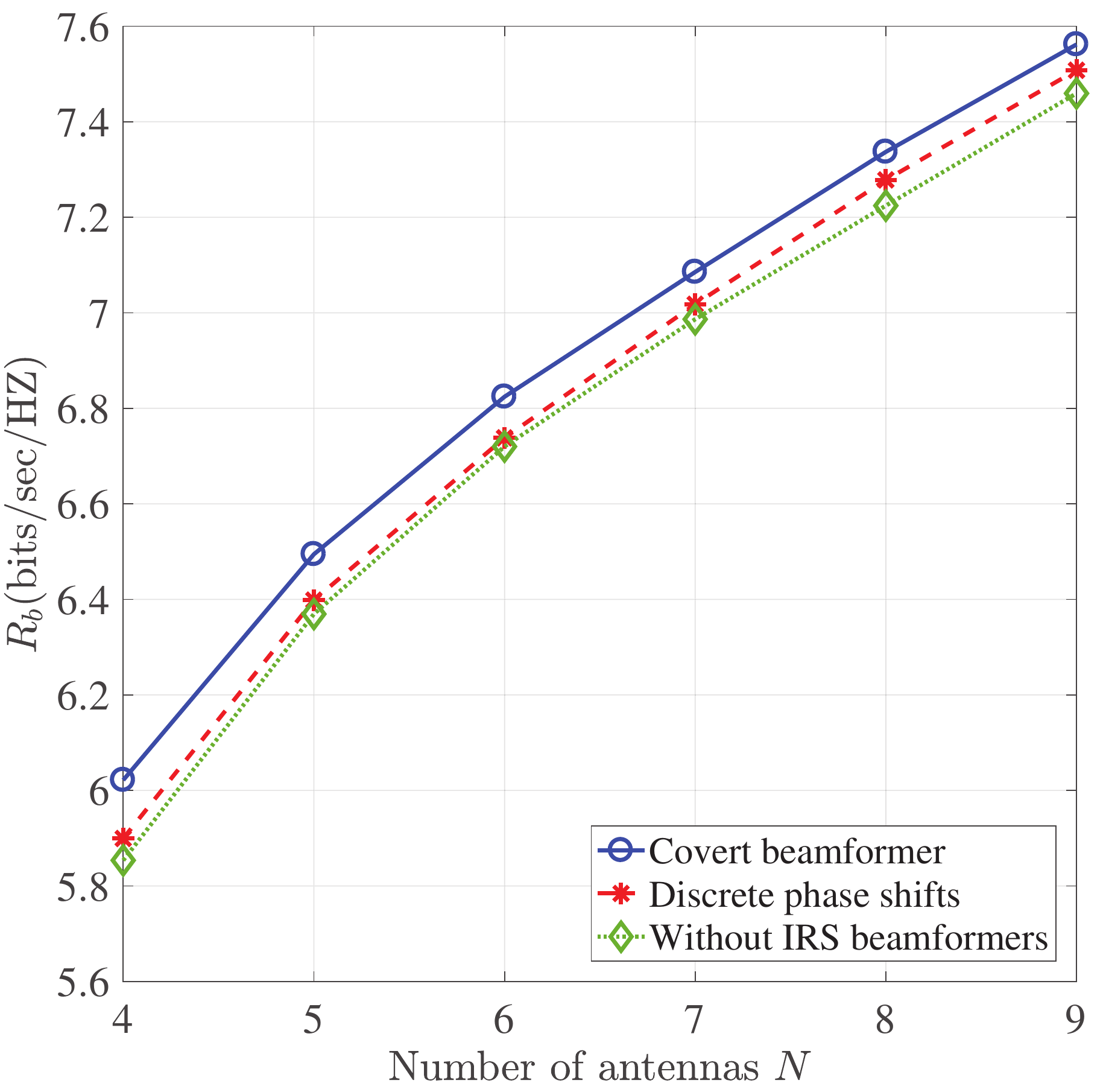}
 \caption{ $R_{\rm{b}}$ versus the number of  antennas $N$.}
  \label{N_rb} 
\end{figure}

In Fig.  \ref{N_rb}, we plot the covert rate of Bob $R_{\rm{b}}$ versus the   number of antennas of Alice $N$    with ${P_{{\rm{total}}}}=-10 \rm{dBm}$ and $K$=4. It can be observed that for a fixed value of $N$,
 $R_{\rm{b}}$ of without IRS beamformer design is lower than that of the  discrete phase shifts design,  while $R_{\rm{b}}$ of the  discrete phase shifts design is lower than that of the covert beamformer design, which is consistent with Fig. \ref{Ptotal_rb}.
Moreover, it can be seen that as the number of antennas $N$ increases, the covert rate of Bob $R_{\rm{b}}$ increases. This is because with more  antennas,  more   spatial multiplexing gains can be exploited.

\subsection{Evaluation for Scenario 2}
In this subsection, the proposed robust beamformer design for scenario that Alice with imperfect WCSI is evaluated.
\begin{figure}
    \begin{minipage}[b]{0.45\textwidth}
      \centering
      \includegraphics[height=8cm,width=7.6cm]{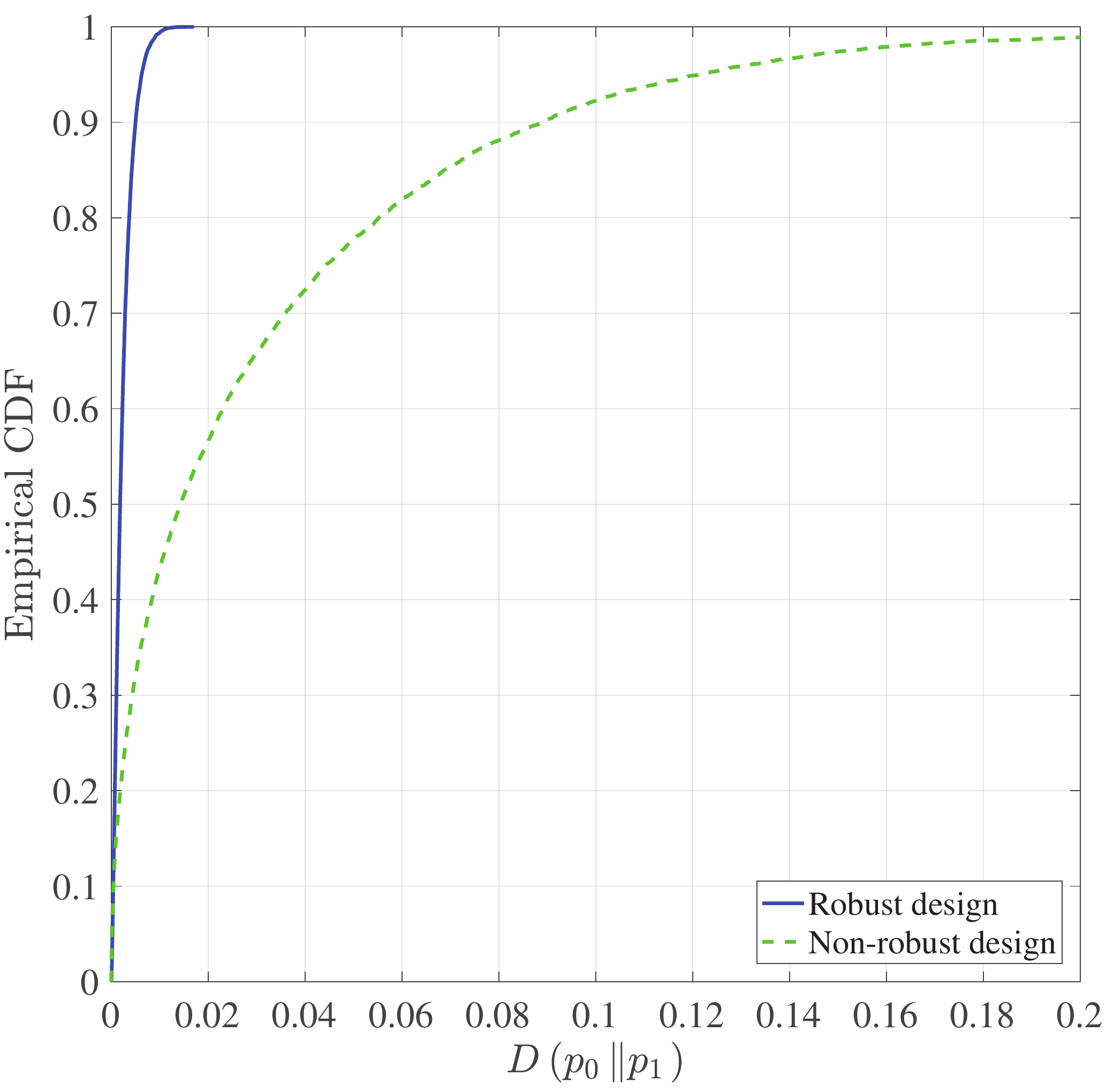}
      \vskip-0.2cm\centering {\footnotesize (a)}
    \end{minipage}
     \begin{minipage}[b]{0.45\textwidth}
      \centering
      \includegraphics[height=8cm,width=7.6cm]{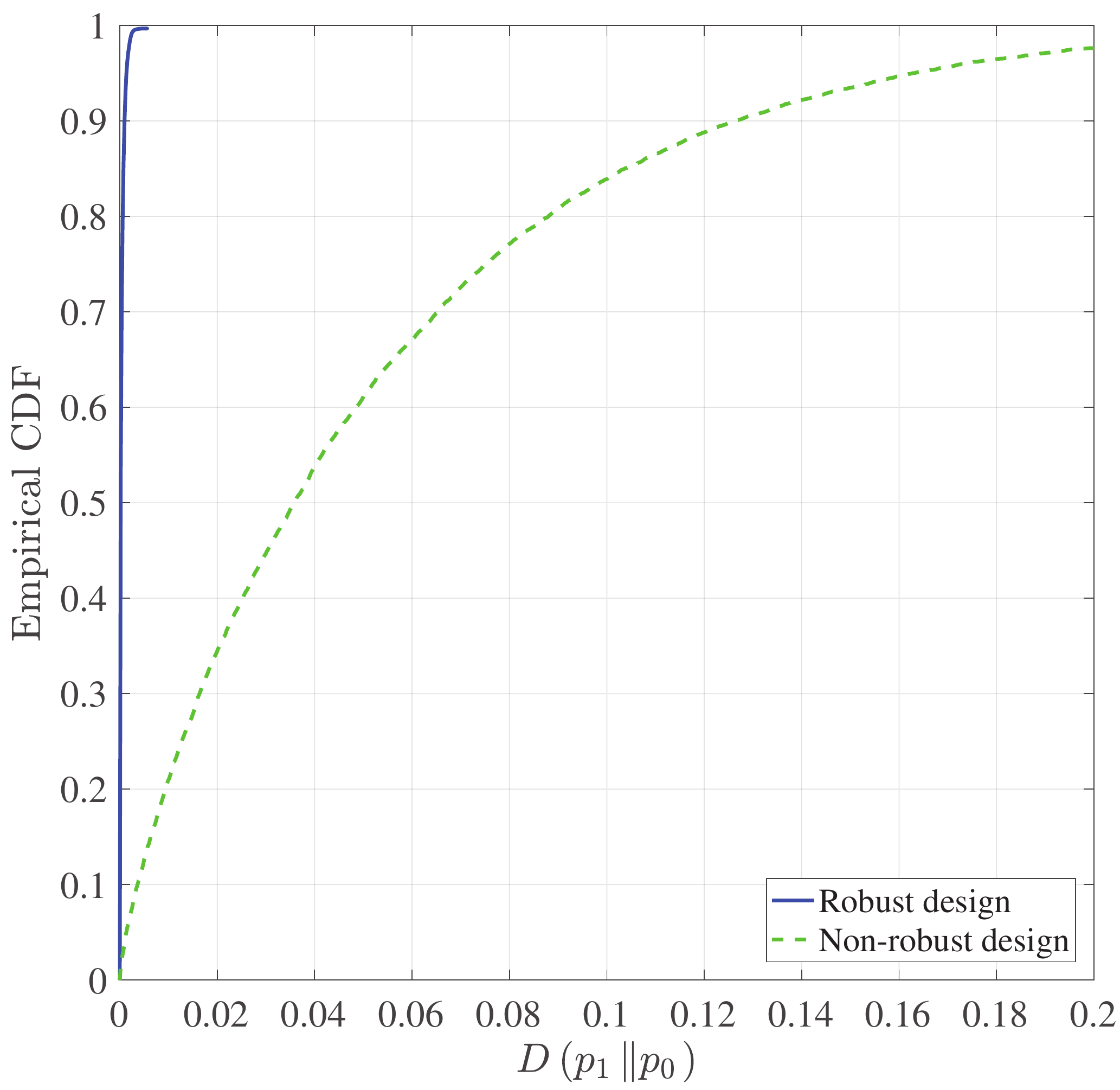}
      \vskip-0.2cm\centering {\footnotesize (b)}
    \end{minipage}\hfill
 \caption{The empirical CDF of (a) $D\left( {{p_0}\left\| {{p_1}} \right.} \right)$  and (b) $D\left( {{p_0}\left\| {{p_1}} \right.} \right)$, with the covertness threshold   $2{\varepsilon ^2} = 0.02$ and  CSI error   $v_w=0.0002$.}
 \label{cdf}  
\end{figure}

Fig. \ref{cdf} (a) and (b) respectively show the empirical  CDF   of the achieved $D\left( {{p_0}\left\| {{p_1}} \right.} \right)$ and $D\left( {{p_1}\left\| {{p_0}} \right.} \right)$ when the CSI error parameter is $v_w={2} \times {{10}}^{{- 4}}$. Here, the non-robust design refers to the proposed covert design with ${{{\bf{\hat h}}}_{\rm{IW}}} $ and ${{{\bf{\hat h}}}_{\rm{AW}}} $ under the same conditions.
For both the robust and non-robust designs, the covertness threshold is $2{\varepsilon ^2} = 0.02$, i.e., $D\left( {{p_0}\left\| {{p_1}} \right.} \right) \le 0.02$ and $D\left( {{p_1}\left\| {{p_0}} \right.} \right) \le 0.02$.
From the Fig. \ref{cdf}, the CDF in the KL divergence of the non-robust design does not satisfy the covert  constraint. In addition, the robust beamforming design guarantees the requirement of the KL divergence. That is, it satisfies Willie's error detection probability requirement.
 In general, Fig. \ref{cdf} (a) and (b) verify     the necessity and effectiveness of the proposed robust design.

\begin{figure}
    \begin{minipage}[b]{0.45\textwidth}
      \centering
      \includegraphics[height=8cm,width=7.6cm]{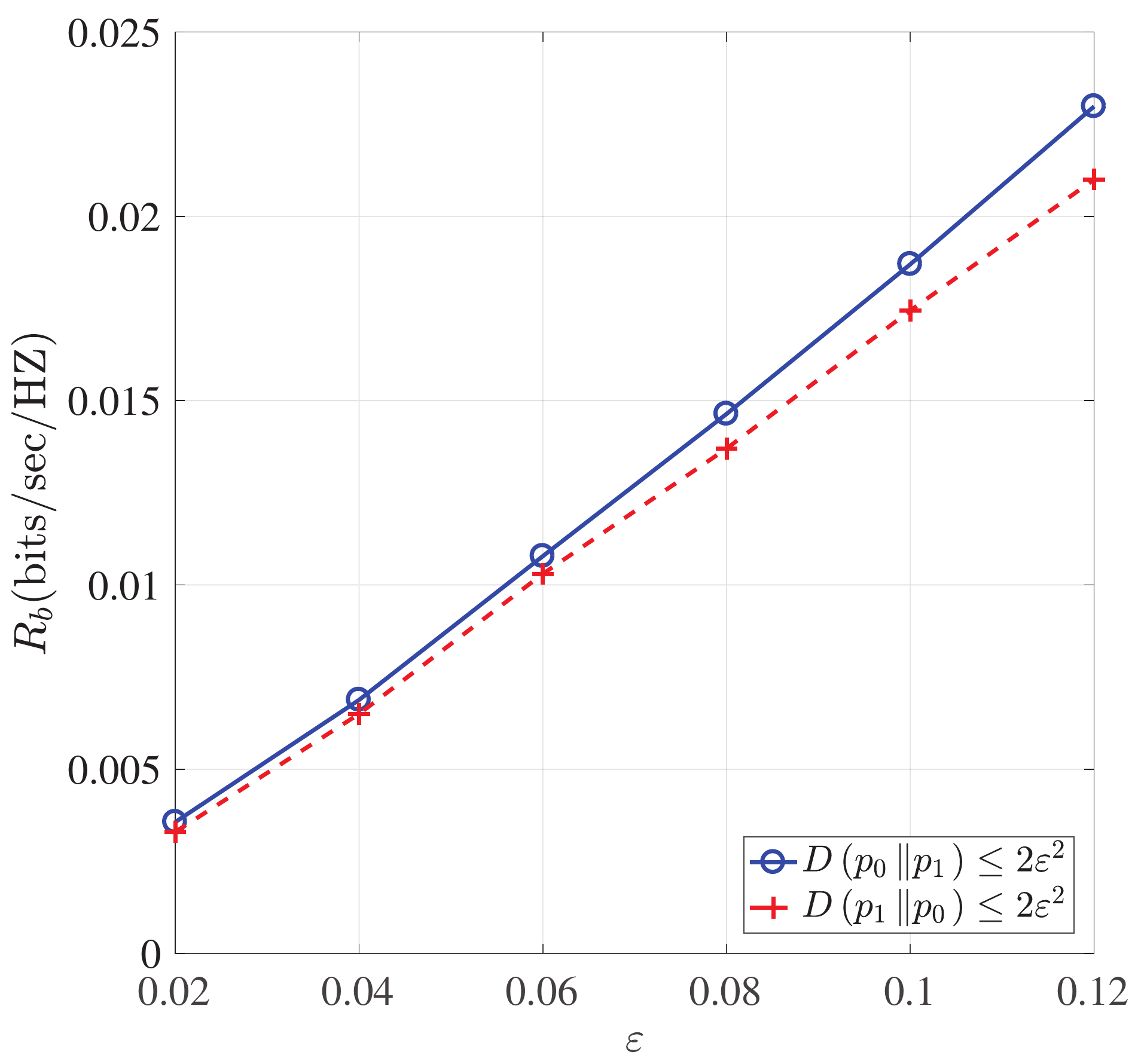}
      \vskip-0.2cm\centering {\footnotesize (a)}
    \end{minipage}
     \begin{minipage}[b]{0.45\textwidth}
      \centering
      \includegraphics[height=8cm,width=8.6cm]{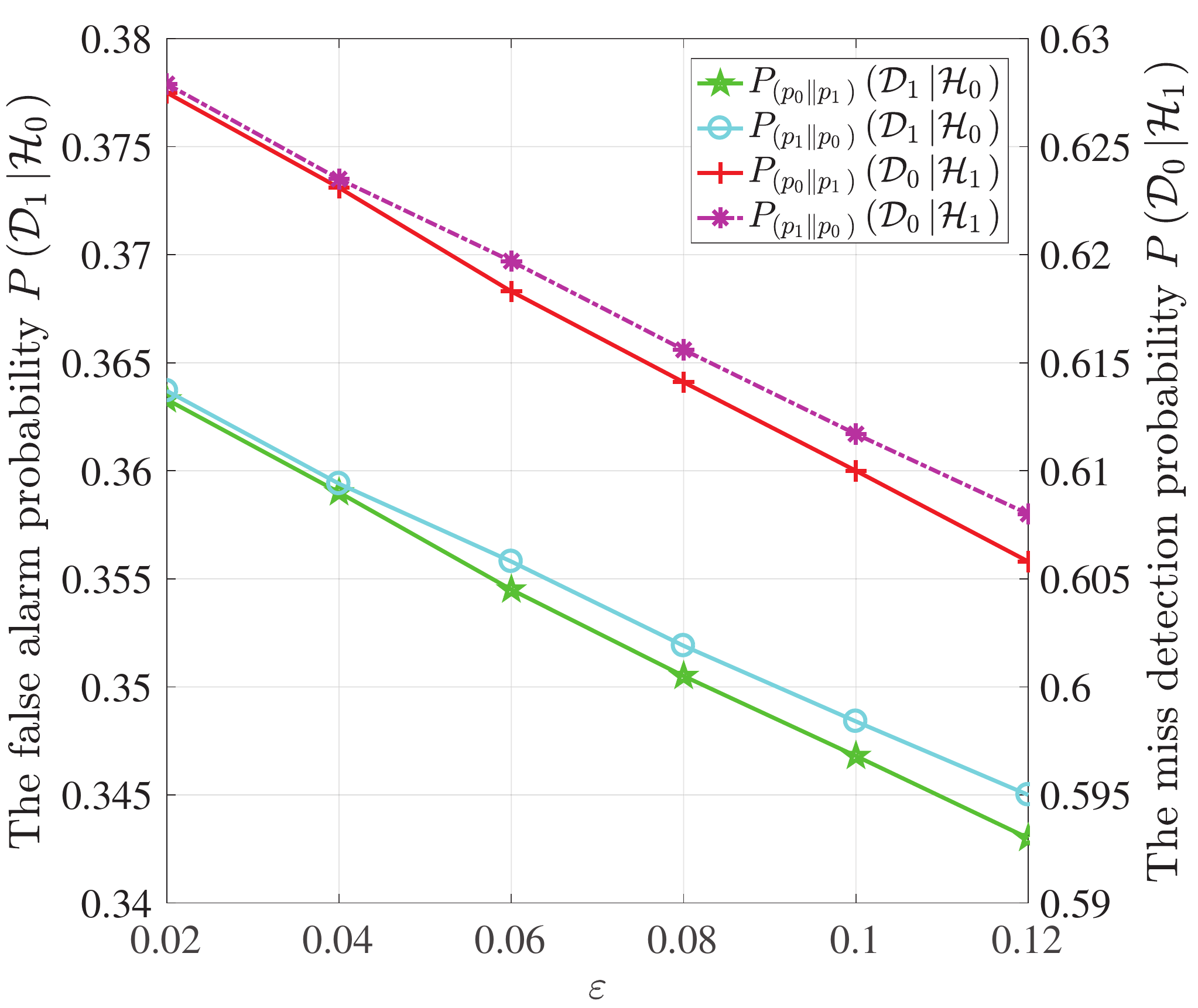}
      \vskip-0.2cm\centering {\footnotesize (b)}
    \end{minipage}\hfill
 \caption{The value of $\varepsilon$ versus (a) the covert rate and (b) the detection error probabilities  with CSI error   $v_w={2} \times {10}^{- 4}$.}
 \label{fig5}  
\end{figure}
\begin{figure}
    \begin{minipage}[b]{0.45\textwidth}
      \centering
      \includegraphics[height=8cm,width=7.6cm]{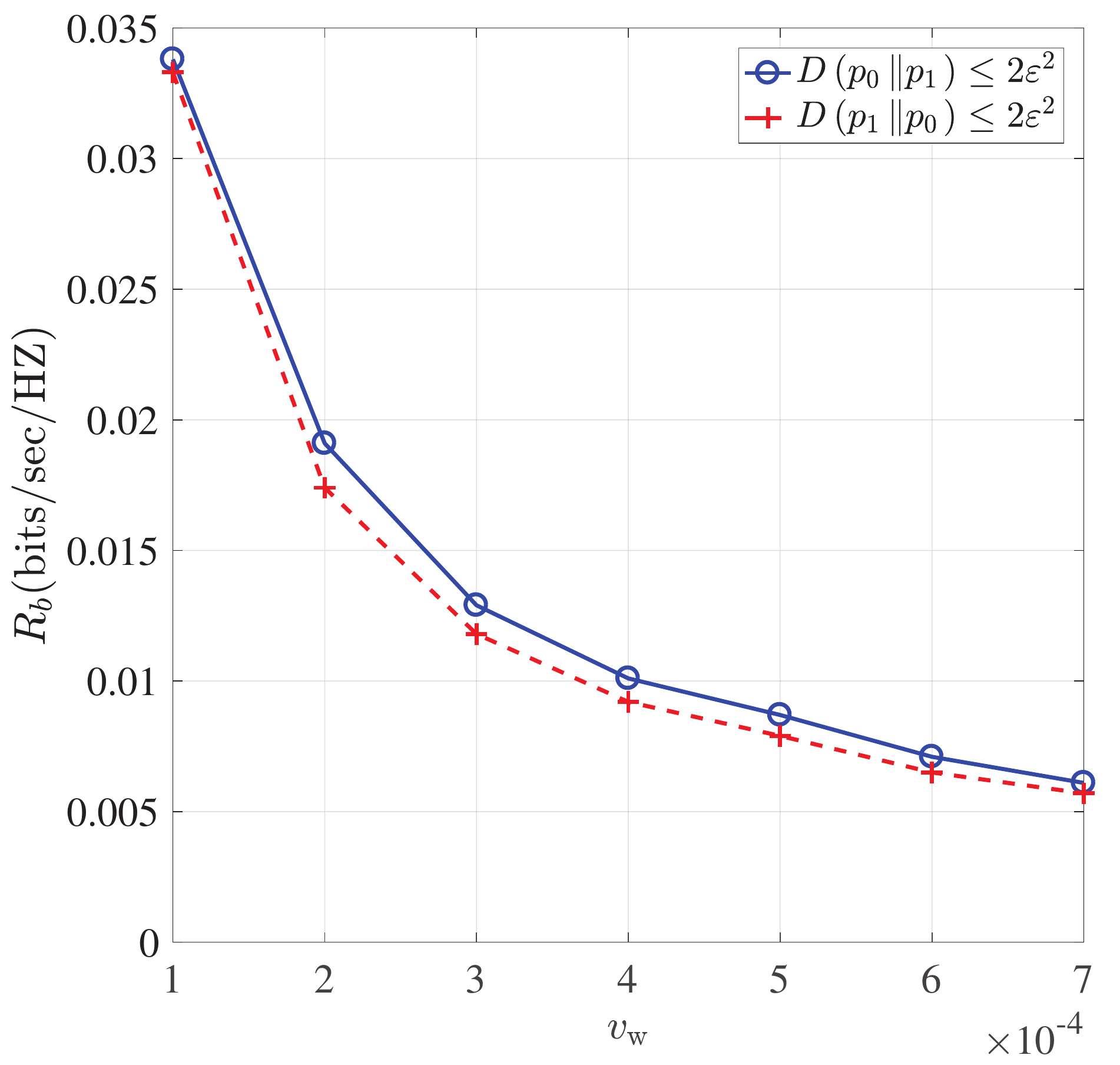}
      \vskip-0.2cm\centering {\footnotesize (a)}
    \end{minipage}
     \begin{minipage}[b]{0.45\textwidth}
      \centering
      \includegraphics[height=8cm,width=8.6cm]{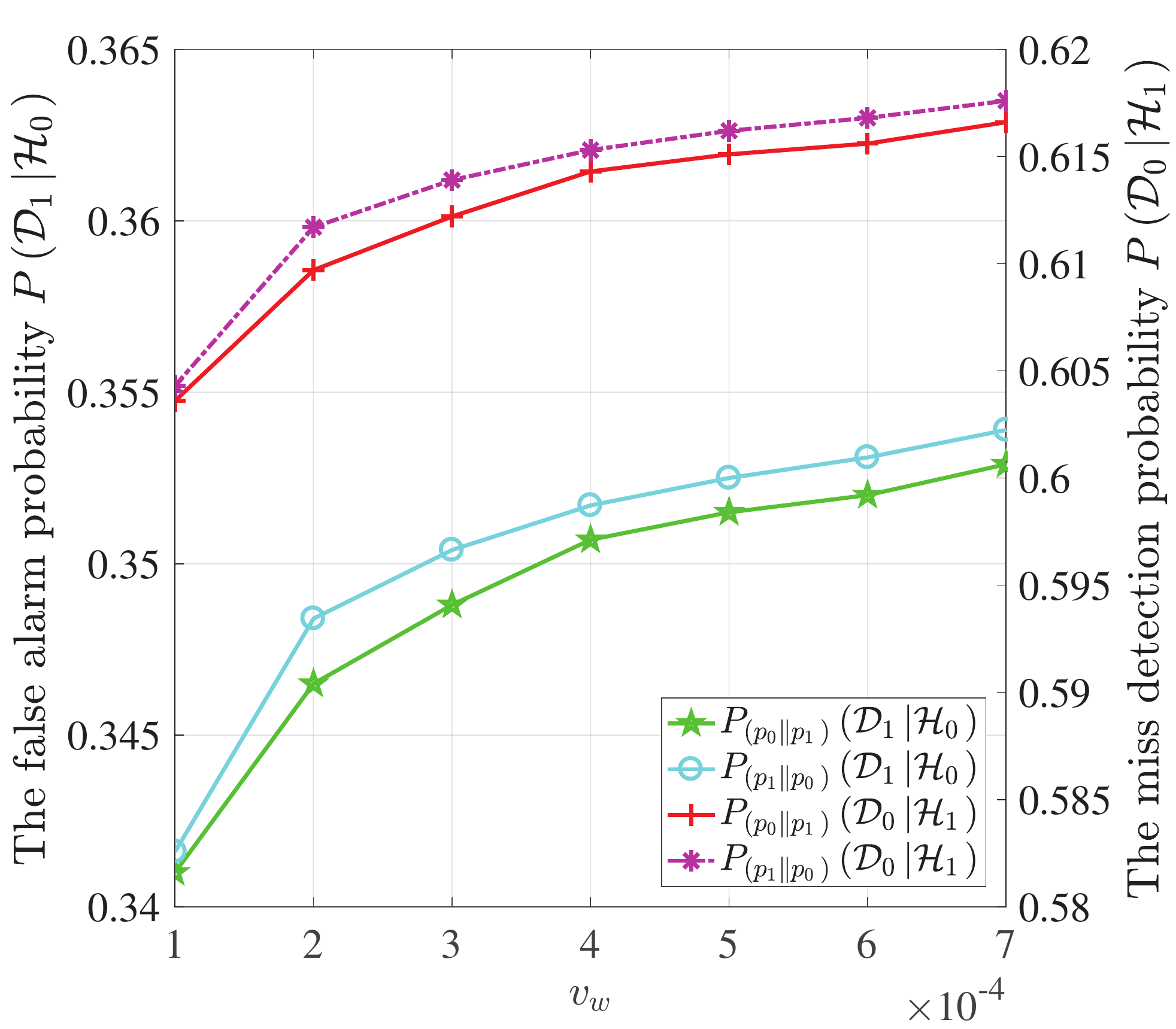}
      \vskip-0.2cm\centering {\footnotesize (b)}
    \end{minipage}\hfill
 \caption{(a) The covert rate and (b) the detection error probabilities
versus CSI error $v_w$ with the value of $\varepsilon=0.1$.
  }
 \label{fig6}  
\end{figure}

 Fig. \ref{fig5} (a) plots the covert rate $R_{\rm{b}}$ versus the value of $\varepsilon$  under
 two covertness constraints $D\left( {{p_0}\left\| {{p_1}} \right.} \right) \le 2{\varepsilon ^2}$ and $D\left( {{p_1}\left\| {{p_0}} \right.} \right) \le 2{\varepsilon ^2}$,
  where CSI error   $v_w={2} \times {{10}}^{{- 4}}$ and ${P_{{\rm{total}}}}=5 \rm{dBm}$. Here, ${P_{\left( {{p_0}\left\| {{p_1}} \right.}\right)}} \left( {{{\cal D}_1}\left| {{{\cal H}_0}} \right.} \right)$ represents the false alarm probability $P \left( {{{\cal D}_1}\left| {{{\cal H}_0}} \right.} \right)$ in the case of $D\left( {{p_0}\left\| {{p_1}} \right.} \right) \le 2{\varepsilon ^2}$, and the other notation is defined likewise.
  This simulation result is consisted with the theoretical analysis. That is, when $\varepsilon$ becomes larger, the covertness constraint becomes loose, which leads to a larger $R_{\rm{b}}$.
   Fig. \ref{fig5} (b) shows that the false alarm probability $P \left( {{{\cal D}_1}\left| {{{\cal H}_0}} \right.} \right)$ and the missed detection probability $P \left( {{{\cal D}_0}\left| {{{\cal H}_1}} \right.} \right)$ versus the value of $\varepsilon$ for CSI error   $v_w={2} \times {{10}}^{{- 4}}$.
 We observe that  under both two different covert constraints,   $P \left( {{{\cal D}_1}\left| {{{\cal H}_0}} \right.} \right)$ and  $P \left( {{{\cal D}_0}\left| {{{\cal H}_1}} \right.} \right)$ are decreasing as    $\varepsilon$ increases, where $P \left( {{{\cal D}_1}\left| {{{\cal H}_0}} \right.} \right)$ is always lower than $P \left( {{{\cal D}_0}\left| {{{\cal H}_1}} \right.} \right)$.
It implies that when the convert constraint is looser, the detection performance of Willie becomes better.
  Moreover,   Fig. \ref{fig5} (b) also verifies    the effectiveness of the proposed robust beamformers design in covert communications, that is,
 $ \Pr \left( {{{\cal D}_1}\left| {{{\cal H}_0}} \right.} \right) + \Pr \left( {{{\cal D}_0}\left| {{{\cal H}_1}} \right.} \right) \ge 1 - \varepsilon$.
 Therefore, from Fig. \ref{fig5}, we reveal the tradeoff between  detection performance of Willie and  covert rate of Bob, and a desired tradeoff can be achieved through an appropriate robust beamforming design.

\begin{figure}[h]
      \centering
	\includegraphics[width=8cm]{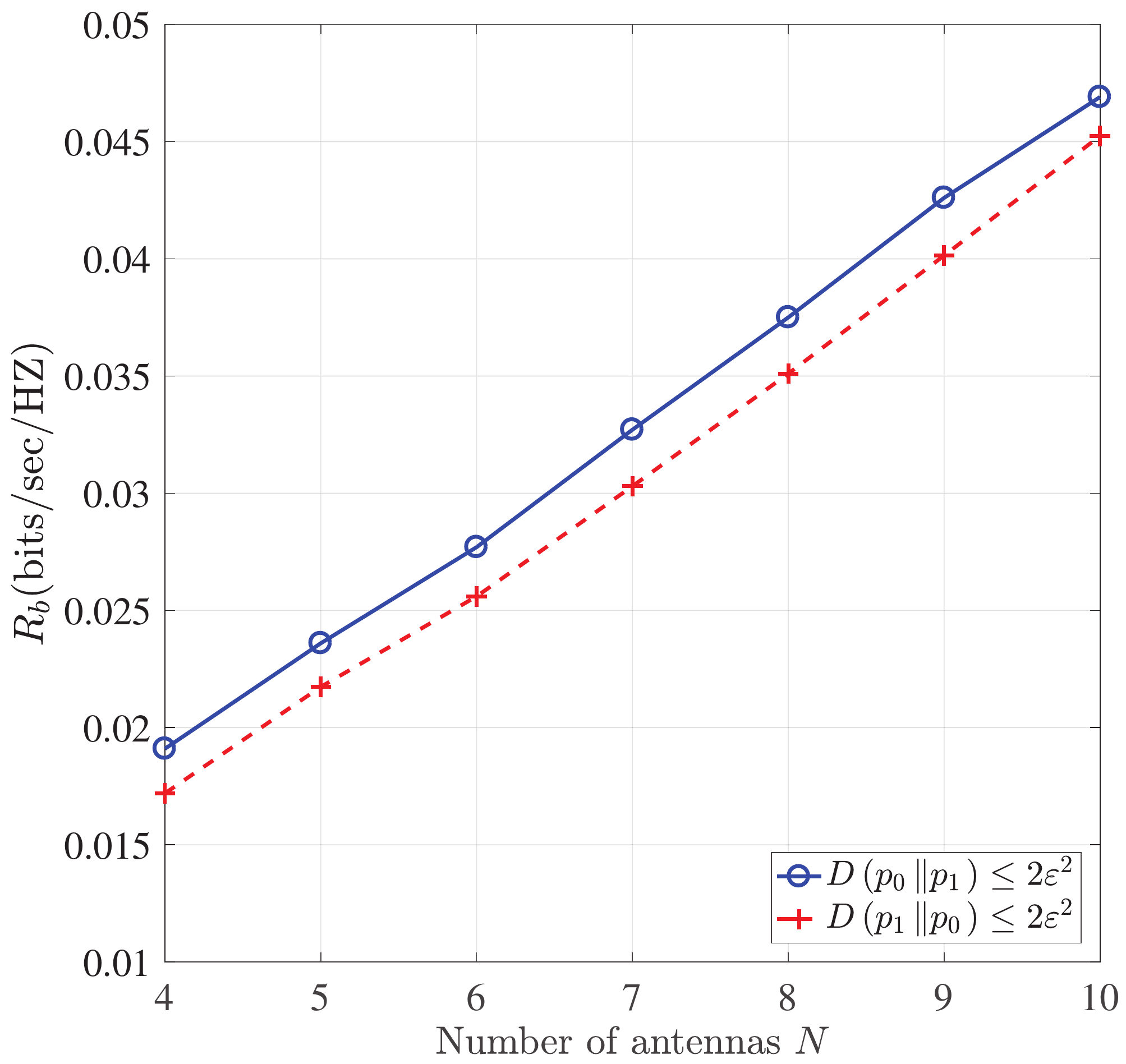}
 \caption{ Covert rates $R_{\rm{b}}$ versus  number of  antennas $N$  with CSI error   $v_w={2} \times {10}^{- 4}$.}
  \label{2_N_rb} 
\end{figure}

 Fig. \ref{fig6} (a) shows  the covert rates $R_{\rm{b}}$ versus CSI error $v_w$ for the two KL divergence cases $D\left( {{p_0}\left\| {{p_1}} \right.} \right) \le 2{\varepsilon ^2}$ and $D\left( {{p_1}\left\| {{p_0}} \right.} \right) \le 2{\varepsilon ^2}$.
 From  Fig. \ref{fig6} (a), we can see that the higher the CSI error $v_w$ is, the lower the achieved covert rates $R_{\rm{b}}$ will be.
    Fig. \ref{fig6} (b) shows the false alarm probability $P \left( {{{\cal D}_1}\left| {{{\cal H}_0}} \right.} \right)$ and the missed detection probability $P \left( {{{\cal D}_0}\left| {{{\cal H}_1}} \right.} \right)$ versus   $v_w$ under
 two covertness constraints.
 We observe that  under the two cases of covertness constraint, both the false alarm probability $P \left( {{{\cal D}_1}\left| {{{\cal H}_0}} \right.} \right)$ and the missed detection probability $P \left( {{{\cal D}_0}\left| {{{\cal H}_1}} \right.} \right)$ decrease when    $v_w$ decreases, where $P \left( {{{\cal D}_1}\left| {{{\cal H}_0}} \right.} \right)$ is always lower than $P \left( {{{\cal D}_0}\left| {{{\cal H}_1}} \right.} \right)$.
 Moreover, Fig. \ref{fig6} shows that a large error $v_w$  may cause a poor beamformer design in terms  of cover rate $R_{\rm{b}}$. However, such beamformer may confuse Willie's detection, which is also beneficial to Bob. Therefore, such a tradeoff also should be paid attention to the  robust beamformer design.

 Fig. \ref{2_N_rb} plots   the covert rates $R_{\rm{b}}$ versus the number of  antennas $N$ under
 two covertness constraints with ${P_{{\rm{total}}}}=5 \rm{dBm}$, $\varepsilon=0.1$ and $v_w={2} \times {10}^{- 4}$.
It can be  observed  that as  the number of antennas $N$ increase, the  covert rates $R_{\rm{b}}$ of  two covertness constraints increase, which is similar to the case in Fig. \ref{2_N_rb}.
  From Fig. \ref{fig5}-\ref{2_N_rb}, we can see that  the rates with the covertness constraint $D\left( {{p_0}\left\| {{p_1}} \right.} \right) \le 2{\varepsilon ^2}$
 are higher than those with   two KL divergence cases  $D\left( {{p_1}\left\| {{p_0}} \right.} \right) \le 2{\varepsilon ^2}$. This is because
    $D\left( {{p_1}\left\| {{p_0}} \right.} \right) \le 2{\varepsilon ^2}$ is stricter than
   $D\left( {{p_0}\left\| {{p_1}} \right.} \right) \le 2{\varepsilon ^2}$, and  this conclusion can also be found in \cite{Yan2019Gaussian}.

\section{Conclusions}

In this paper, we designed  both covert beamformer   and robust beamformer    for IRS assisted IoT networks, where Alice utilizes the IRS to covertly send a message to Bob to avoid being discovered by Willie.
  For the   perfect WCSI scenario, we develop  the  covert beamformer design for the    covert rate maximization, and the numerical results show that the  covert beamformer design has better covert performance than the design without IRS.
Furthermore, for practical   imperfect WCSI scenario,
 we derived
 the covert decision threshold of Willie, the false alarm
probability, and the missed detection probability.
Then,
 by taking the impact of practical channel estimation errors into account,   we  proposed   robust beamformers design,
which can maximize the covert rate   while meeting the covert requirements.
 Numerical results illustrated the validity of the proposed covert  beamformers design and
provided the useful insights on the effect of the main   design  parameters
on the covert communication   performance.

  \begin{appendices}

\section{{Proof of the Proposition 1:}}

Let ${\bf{P}}$ denotes the projection matrix of vector ${{\bf{W}}_{\rm{b}}^{{1 \mathord{\left/
 {\vphantom {1 2}} \right.
 \kern-\nulldelimiterspace} 2}}{\bf{t}}_{\rm{b}}^H}$, where  ${\bf{W}}_{\rm{b}}$ is an SDR solution for problem \eqref{problem1B},
 \begin{align}
 {\bf{P}} = \frac{{{\bf{W}}_{\rm{b}}^{{1 \mathord{\left/
 {\vphantom {1 2}} \right.
 \kern-\nulldelimiterspace} 2}}{\bf{t}}_{\rm{b}}^H{{\left( {{\bf{W}}_{\rm{b}}^{{1 \mathord{\left/
 {\vphantom {1 2}} \right.
 \kern-\nulldelimiterspace} 2}}{\bf{t}}_{\rm{b}}^H} \right)}^H}}}{{{{\left\| {{\bf{t}}_{\rm{b}}^H{\bf{W}}_{\rm{b}}^{{1 \mathord{\left/
 {\vphantom {1 2}} \right.
 \kern-\nulldelimiterspace} 2}}} \right\|}^2}}} = \frac{{{\bf{W}}_{\rm{b}}^{{1 \mathord{\left/
 {\vphantom {1 2}} \right.
 \kern-\nulldelimiterspace} 2}}{\bf{t}}_{\rm{b}}^H{{\bf{t}}_{\rm{b}}}{\bf{W}}_{\rm{b}}^{{1 \mathord{\left/
 {\vphantom {1 2}} \right.
 \kern-\nulldelimiterspace} 2}}}}{{{{\left( {{\bf{W}}_{\rm{b}}^{{1 \mathord{\left/
 {\vphantom {1 2}} \right.
 \kern-\nulldelimiterspace} 2}}{\bf{t}}_{\rm{b}}^H} \right)}^H}{\bf{W}}_{\rm{b}}^{{1 \mathord{\left/
 {\vphantom {1 2}} \right.
 \kern-\nulldelimiterspace} 2}}{\bf{t}}_{\rm{b}}^H}},
 \end{align}

 We construct a new rank one solution ${{{\bf{\bar W}}}_{\rm{b}}} = {\bf{W}}_{\rm{b}}^{{1 \mathord{\left/
 {\vphantom {1 2}} \right.
 \kern-\nulldelimiterspace} 2}}{\bf{PW}}_{\rm{b}}^{{1 \mathord{\left/
 {\vphantom {1 2}} \right.
 \kern-\nulldelimiterspace} 2}}$. Then, let us check the value of the
objective function ${{\bf{W}}_{\rm{b}}} - {{{\bf{\bar W}}}_{\rm{b}}}{\rm{ = }}{\bf{W}}_{\rm{b}}^{{1 \mathord{\left/
 {\vphantom {1 2}} \right.
 \kern-\nulldelimiterspace} 2}}\left( {{\bf{I}} - {\bf{P}}} \right){\bf{W}}_{\rm{b}}^{{1 \mathord{\left/
 {\vphantom {1 2}} \right.
 \kern-\nulldelimiterspace} 2}}\underline  \succ  {\bf{0}}$. Thus,
${\rm{Tr}}\left( {{{{\bf{\bar W}}}_{\rm{b}}}} \right) \le {\rm{Tr}}\left( {{{\bf{W}}_{\rm{b}}}} \right) \le {P_{{\rm{total}}}}$, which means the solution ${{{\bf{\bar W}}}_{\rm{b}}} $ satisfies constraint \eqref{problem1c}. Moreover, substituting ${{{\bf{\bar W}}}_{\rm{b}}} $ into the value of the objective function, we have
 \begin{align}
{{\bf{t}}_{\rm{b}}}{{{\bf{\bar W}}}_{\rm{b}}}{\bf{t}}_{\rm{b}}^H& = {{\bf{t}}_{\rm{b}}}{\bf{W}}_{\rm{b}}^{{1 \mathord{\left/
 {\vphantom {1 2}} \right.
 \kern-\nulldelimiterspace} 2}}{\bf{PW}}_{\rm{b}}^{{1 \mathord{\left/
 {\vphantom {1 2}} \right.
 \kern-\nulldelimiterspace} 2}}{\bf{t}}_{\rm{b}}^H\nonumber\\
 &= \frac{{{{\bf{t}}_{\rm{b}}}{\bf{W}}_{\rm{b}}^{{1 \mathord{\left/
 {\vphantom {1 2}} \right.
 \kern-\nulldelimiterspace} 2}}{\bf{W}}_{\rm{b}}^{{1 \mathord{\left/
 {\vphantom {1 2}} \right.
 \kern-\nulldelimiterspace} 2}}{\bf{t}}_{\rm{b}}^H{{\bf{t}}_{\rm{b}}}{\bf{W}}_{\rm{b}}^{{1 \mathord{\left/
 {\vphantom {1 2}} \right.
 \kern-\nulldelimiterspace} 2}}{\bf{W}}_{\rm{b}}^{{1 \mathord{\left/
 {\vphantom {1 2}} \right.
 \kern-\nulldelimiterspace} 2}}{\bf{t}}_{\rm{b}}^H}}{{{{\left( {{\bf{W}}_{\rm{b}}^{{1 \mathord{\left/
 {\vphantom {1 2}} \right.
 \kern-\nulldelimiterspace} 2}}{\bf{t}}_{\rm{b}}^H} \right)}^H}{\bf{W}}_{\rm{b}}^{{1 \mathord{\left/
 {\vphantom {1 2}} \right.
 \kern-\nulldelimiterspace} 2}}{\bf{t}}_{\rm{b}}^H}}\nonumber\\
 &= {{\bf{t}}_{\rm{b}}}{{\bf{W}}_{\rm{b}}}{\bf{t}}_{\rm{b}}^H.
 \end{align}
Hence, the value of the objective function remains the same ${\bf{W}}_{\rm{b}}$ is replaced
with ${{{\bf{\bar W}}}_{\rm{b}}} $. Finally, let us check whether the constraint \eqref{problem1Bb} is satisfied for the new solution ${{{\bf{\bar W}}}_{\rm{b}}} $
 \begin{align}
{{\bf{t}}_{\rm{w}}}{{\bf{W}}_{\rm{b}}}{\bf{t}}_{\rm{w}}^H - {{\bf{t}}_{\rm{w}}}{{{\bf{\bar W}}}_{\rm{b}}}{\bf{t}}_{\rm{w}}^H  &= {{\bf{t}}_{\rm{w}}}\left( {{{\bf{W}}_{\rm{b}}} - {\bf{W}}_{\rm{b}}^{{1 \mathord{\left/
 {\vphantom {1 2}} \right.
 \kern-\nulldelimiterspace} 2}}{\bf{PW}}_{\rm{b}}^{{1 \mathord{\left/
 {\vphantom {1 2}} \right.
 \kern-\nulldelimiterspace} 2}}} \right){\bf{t}}_{\rm{w}}^H\nonumber\\
  &= {{\bf{t}}_{\rm{w}}}\left( {{\bf{W}}_{\rm{b}}^{{1 \mathord{\left/
 {\vphantom {1 2}} \right.
 \kern-\nulldelimiterspace} 2}}\left( {{\bf{I}} - {\bf{P}}} \right){\bf{W}}_{\rm{b}}^{{1 \mathord{\left/
 {\vphantom {1 2}} \right.
 \kern-\nulldelimiterspace} 2}}} \right){\bf{t}}_{\rm{w}}^H\nonumber\\
 &\ge {\rm{0}}.
 \end{align}
Due to the value of the ${\rm{Tr}}\left( {{{\bf{t}}_{\rm{w}}}{{\bf{W}}_{\rm{b}}}{\bf{t}}_{\rm{w}}^H} \right){\rm{ = 0}}$, thus, ${\rm {Tr}}\left( {{{\bf{t}}_{\rm{w}}}{{{\bf{\bar W}}}_{\rm{b}}}{\bf{t}}_{\rm{w}}^H} \right){\rm{ = 0}}$.

\end{appendices}

%
%
\bibliographystyle{IEEE-unsorted}
\bibliography{refs0611}

\end{document}